\documentclass[fleqn,usenatbib]{mnras}

\usepackage{newtxtext,newtxmath}
\usepackage[T1]{fontenc}

\DeclareRobustCommand{\VAN}[3]{#2}
\let\VANthebibliography\thebibliography
\def\thebibliography{\DeclareRobustCommand{\VAN}[3]{##3}\VANthebibliography}

\usepackage{graphicx}

\usepackage{amsmath}	% Advanced maths commands
\newcommand\massr{M}
\newcommand\massh{m}
\newcommand\mapp{m_{\rm AB}}
\newcommand\mabs{M_{\rm AB}}
\newcommand\pnum{2000~} 
\newcommand\regeff{R_{\rm eff}}
\newcommand\eulrad{R_{\rm e}}
\newcommand\lagrad{R}
\newcommand\Full{\textit{Full}}
\newcommand\NoCV{\textit{NoCV}}
\newcommand\Standard{\textit{Standard}}
\newcommand\Incorrect{\textit{Naive}}
\newcommand\pakidge{\textbf{galcv}}

\title[Cosmic Variance]{A Flexible Analytic Model of Cosmic Variance in the First Billion Years}

\author[Trapp \& Furlanetto]{
A.C. Trapp,$^{1}$\thanks{E-mail: atrapp@astro.ucla.edu}
Steven R. Furlanetto,$^{1}$
\\
$^{1}$Department of Physics and Astronomy, University of California Los Angeles, CA, 90095-1562, USA \\
}

\date{Accepted 2020 September 9. Received 2020 September 9; in original form 2020 August 14}

\pubyear{2020}

\begin{document}
\label{firstpage}
\pagerange{\pageref{firstpage}--\pageref{lastpage}}
\maketitle

\begin{abstract}
Cosmic variance is the intrinsic scatter in the number density of galaxies due to fluctuations in the large-scale dark matter density field. In this work, we present a simple analytic model of cosmic variance in the high redshift Universe ($z\sim5$--15). 
We assume that galaxies grow according to the evolution of the halo mass function, which we allow to vary with large-scale environment. 
Our model produces a reasonable match to the observed ultraviolet luminosity functions in this era by regulating star formation through stellar feedback and assuming that the UV luminosity function is dominated by recent star formation. 
We find that cosmic variance in the UVLF is dominated by the variance in the underlying dark matter halo population, and not by differences in halo accretion or the specifics of our stellar feedback model. 
We also find that cosmic variance dominates over Poisson noise for future high-$z$ surveys except for the brightest sources or at very high redshifts ($z \gtrsim 12$).
We provide a linear approximation of cosmic variance for a variety of redshifts, magnitudes, and survey areas through the public Python package \pakidge. 
Finally, we introduce a new method for  incorporating priors on cosmic variance into estimates of the galaxy luminosity function and demonstrate that it significantly improves constraints on that important observable.
\end{abstract}


%%%%%%%%%%%%%%%%%%%%%%%%%%%%%%%%%%%%%%%%%%%%%%%%%%

%%%%%%%%%%%%%%%%% BODY OF PAPER %%%%%%%%%%%%%%%%%%

\section{Introduction}

Extragalactic astronomy is closing in on arguably the most important era of galaxy evolution: the formation of the first galaxies. These galaxies will allow us to probe the processes that drove the first emergence of complexity in our Universe.

The most fundamental observations for studying this era have been (and will continue to be) deep galaxy surveys. These surveys measure many important features of the galaxy population, most fundamentally the ultra-violet luminosity function (UVLF) of galaxies. The UVLF is a measure of the number of galaxies at each luminosity, and its shape and evolution through cosmic time has important implications for the physics behind galaxy formation and growth, and much more \citep[see e.g.,][]{Bouwens2015, Finkelstein2015, Livermore2017, Atek2018, Oesch2018, Behroozi2019}. Unfortunately, these deep galaxy surveys will have very small volumes, which will be a key limitation in measuring the UVLF due to the effects of ``cosmic variance:'' not all regions of the Universe contain the average number of galaxies, and those galaxies did not all grow up in an average environment. We must understand how cosmic variance affects the UVLF in order to inform and correctly interpret future deep galaxy surveys. 

Cosmic variance in the UVLF (and other measures) has been modeled in a variety of ways in the past. For example, analytic models typically start with the linear halo bias function and then connect haloes to galaxies with a halo mass to luminosity relation, or by matching abundances \citep[see e.g.,][]{Newman2002,Somerville2004,Stark2007,Moster2011}. These models conclude that cosmic variance is a significant source of uncertainty when studying galaxies at high redshifts. However, such models do not allow cosmic variance to affect the halo mass to luminosity connection itself; they assume galaxies are the same in all environments. Also, the linear halo bias function does not accurately predict cosmic variance in extreme environments.

Cosmic variance can also be estimated using mock observations of galaxy simulations. The early implementations of this method \citep[see e.g.,][]{Kitzbichler2007,Trenti2008} were very powerful, but also assumed galaxies are the same in all environments. Recently, substantial improvements in computing power have allowed for much higher-volume $N$-body simulations that also treat star formation in a more complex way \citep[e.g.,][]{Bhowmick2020,Ucci2020}. These studies take into account the difference in environment on individual galaxy growth, and are a major step forward in predicting cosmic variance in the first galaxies.
However, these studies (1) are still limited by their volume, as a complete picture of cosmic variance requires extremely large volumes to calculate cosmic variance on all relevant scales and magnitudes; (2) cannot explore how cosmic variance depends on their specific implementation of mass accretion, star formation, feedback, and other parameters, without re-running simulations many times, which would be prohibitively expensive; and (3) can be limited in their redshift or magnitude ranges.

Simulations lack large volume and flexibility, while existing analytic models rely on linear theory and lack a fully self-consistent connection between cosmic variance and galaxy growth and star formation.

Quantitative interpretations of high-$z$ data require corrections for cosmic variance, especially because most planned surveys subtend relatively small volumes. Such corrections are particularly important when multiple independent surveys are combined, as each such survey contains its own (unknown) intrinsic density. The standard method to account for cosmic variance when fitting a UVLF, originally developed by \citet{Sandage1979} and used by e.g. \citet{Efstathiou1988} and \citet{Bouwens2015}, fits a universal \textit{shape} of the UVLF to all fields, ignoring the normalization parameter of the fit in each individual field. After the shape has been optimized, the overall normalization is determined by demanding that it reproduces the correct total number of galaxies across all surveys. This method cannot account for a change in shape of the UVLF between fields, and it does not include a prior for the amount of variance allowed in the normalization parameter of the UVLF.

In this paper, we use a simple, flexible analytic model of high-$z$ galaxies to study the effects of cosmic variance on galaxy surveys. We begin in section \ref{sec:haloes} with a close examination of large-scale variations in the dark matter halo population in the context of excursion set models of halo formation. This provides the fundamental basis for the cosmic variance of the galaxy population, and with it we capture some non-linear aspects of these fluctuations. Next, in section \ref{sec:FRSF}, we describe a ``minimalist" model of galaxy evolution \citep{Furlanetto2017} that fits observed luminosity functions reasonably well but is sufficiently flexible to examine how a large range of assumptions about the physics of these sources affects cosmic variance. We use this model to determine which of the many uncertain parameters of galaxy formation have the most impact on cosmic variance, and we account for changes in galaxy growth and star formation in different environments. We combine our treatments of dark matter haloes and galaxy physics in section \ref{sec:cosvar}, where we also provide a linear approximation to the cosmic variance of galaxies as a function of redshift and absolute magnitude. Unlike other such functions derived from simulations, our results apply across any mass or redshift, and we quantify how uncertainties in galaxy evolution parameters affect the results.

In section \ref{sec:survey}, we then describe the importance of our cosmic variance results for future surveys of high-$z$ galaxies with the James Webb Space Telescope (JWST) and the Nancy Grace Roman Space Telescope \citep[hereafter, the Roman Space Telescope;][]{Spergel2015,Akeson2019,Dore2019}. We show how cosmic variance limits inferences about the average UVLF of the Universe. Additionally, we introduce a method that fully incorporates cosmic variance into UVLF estimates, essentially treating our estimates for cosmic variance as a prior on the measurements. Most commonly, UVLF estimates allow for an arbitrary amount of cosmic variance between fields, by ignoring the normalization of the UVLF in each field (e.g., \citealt{Bouwens2015, Finkelstein2015}, though see \citealt{Livermore2017} for a contrasting case). We show that our method provides tighter constraints in mock surveys. Finally, in section \ref{sec:conclusions}, we summarize our results.

We take the following cosmological parameters: $\Omega_m = 0.308$, $\Omega_\Lambda=0.692$, $\Omega_b=0.0484$, $h=0.678$, $\sigma_8=0.815$, and $n_s=0.968$, consistent with recent Planck Collaboration XIII results \citep{PlanckCollaboration2016}. We give all distances in comoving units.

%%%%%%%%%%%%%%%%%%%%%%%%%%%%%%%%%%%%%%%%%%%%%%%%%%%%%%%%%%%%%%%%
%%%%%%%%%%%%%%%%%%%%%%%%%%%%%%%%%%%%%%%%%%%%%%%%%%%%%%%%%%%%%%%%

%%%%%%%%%%%%%%%%%%%%%%%%%%%%%%%%%%%%%%%%%%%%%%%%%%%%%%%%%%%%%%%%
%%%%%%%%%%%%%%%%%%%%%%%%%%%%%%%%%%%%%%%%%%%%%%%%%%%%%%%%%%%%%%%%
\section{Dark Matter Haloes}\label{sec:haloes}

We follow the methods described in \citet{Furlanetto2017} to model dark matter haloes. In this section we give a brief summary of those methods and also describe some additions.

%%%%%%%%%%%%%%%%%%%%%%%%%%%%%%%%%%%%%%%%%%%%%%%%%%%%%%%%%%%%%%%%
\subsection{Conditional halo mass function}\label{sec:CMF}

We define the dark matter halo mass function as $n_{\rm h}(\massh,z) d\massh$: the comoving number density of dark matter haloes between masses $(\massh,\massh+d\massh)$ at redshift $z$.
By convention,
\begin{equation}\label{eq:massfun}
    n_{\rm h}(\massh,z)=f(\sigma)\frac{\bar\rho}{\massh}\textrm{dln}\frac{(1/\sigma)}{\textrm{d}\massh},
\end{equation}
where $\bar\rho$ is the comoving average matter density, $\sigma(\massh,z)$ is the linear rms fluctuation of the matter density field at redshift $z$ smoothed over a spherical region of mass $\massh$ (see section \ref{sec:pb} for the calculation of $\sigma(m,z)$), and $f(\sigma)$ is a dimensionless function that modifies the shape of the mass function. Following \citet{Furlanetto2017}, we use $f_{\rm Trac}(\sigma)$ from a fit to the average mass function of a high-$z$ cosmological simulation \citep{Trac2015}:
\begin{equation}\label{eq:trac}
    f_{\rm Trac}(\sigma)=0.150\left[ 1+\left(\frac{\sigma}{2.54}\right)^a\right]e^{b/\sigma^2},
\end{equation}
with $a=-1.36$ and $b=-1.14$. 

The key aspect of this model is allowing the halo mass function to depend on its environment. This environmental dependence is introduced via the conditional mass function (CMF).
The CMF $n_{\textrm{\rm cond}}(\massh,z,\delta_b, \lagrad)$ describes the number density of haloes in a spherical region of mass $\massr$, with a corresponding Lagrangian radius\footnote{Note that the radius $\lagrad$ is really a mass scale, as it does not correspond to the real radius of a region except for regions that happen to be at cosmological average density.} $\lagrad^3 = 3\massr/(4\pi\bar{\rho})$, and relative density $\delta_b = (\rho-\bar{\rho})/\bar{\rho}$, where $\rho$ is the linearly extrapolated matter density in that region at redshift $z$.
The CMF is what adds cosmic variance into the model. 

We determine $n_{\textrm{\rm cond}}(\massh,z,\delta_b, \lagrad)$ using a coordinate transfer method described in \citet{Tramonte2017} that we will call ``$\nu$-scaling'' applied to equation~(\ref{eq:trac})\footnote{We construct the CMF this way because simulations of dark matter haloes \citep[e.g.][]{McBride2009,Goerdt2015,Trac2015} do not provide a full CMF.}
\begin{equation}\label{eq:stdscl}
\begin{aligned}
    \sigma^2(\massh,z) &\rightarrow 
    [\sigma^2(\massh,z)-\sigma^2(\massr,z)]\left[ \frac{\delta_{crit}}{\delta_{crit}-\delta_b} \right]^2,
\end{aligned}
\end{equation}
where $\massr$ is the mass corresponding to the region $\lagrad$, and $\delta_{crit}\approx$1.69 is the linear halo collapse threshold \citep[see][eq.(3.13)]{Loeb2013}. The resulting CMF is Lagrangian in that it assumes all regions of fixed mass have the same volume. To convert into a real-space (Eulerian) CMF, we calculate the real-space radius $\eulrad$ of a region of mass $\massr$ and density $\delta_b$ assuming spherical collapse: $\eulrad = \lagrad/(1+\delta_r)^{1/3}$, where $\delta_r$ is the real-space (non-linear) relative density\footnote{In practice, $\delta_r$ and $\delta_b$ are very similar, especially at the redshifts considered in this paper.}\citep[][see Appendix \ref{app:EulCorr} for more details]{Mo1996}. Applying this adjustment to the radius of each region results in an Eulerian CMF, $n_{\rm cond}(m,z,\delta_b,\eulrad) = n_{\rm cond}(m,z,\delta_b,\lagrad)\times(1+\delta_r)$.

\citet{Tramonte2017} justify $\nu$-scaling by noting $\delta_{crit}$ enters into $f(\sigma)$ only through the variable $\nu=\delta_{crit}/\sigma$ in previous parameterizations. They then apply the ``standard'' coordinate transfer
\begin{equation}\label{eq:coordtrans}
\begin{aligned}
    \delta_{crit} &\rightarrow \delta_{crit} - \delta_b\\
    \sigma^2(\massh,z) &\rightarrow \sigma^2(\massh,z) - \sigma^2(\massr,z)
\end{aligned}
\end{equation}
to the variable $\nu$, giving equation~(\ref{eq:stdscl}). \citet{Tramonte2017} validate this method using an N-body simulation by \citet{Tinker2008}, finding that this scaling technique accurately describes the CMF except for the most underdense regions ($\delta_b \lesssim -1.5$), where it overestimates halo abundance.

We note that while we consider halo masses down to $\massh \sim 10^{8} \textrm{M}_\odot$, \citet{Tramonte2017} test their prescription only down to a mass of $\massh \sim 3\times10^{10} \textrm{M}_\odot$, leaving it untested for the lowest masses we consider. We also note that we use a different mass function \citep{Trac2015}, for which this method has not been explicitly tested. \citet{Tramonte2017} also test ``local scaling'', a more rigorous method for constructing a CMF developed in \citet{Patiri2006} and expanded in \citet{RubinoMartin2008}, and find that method produces a slightly better CMF. We do not test our model using local scaling as it cannot be easily applied to the \citet{Trac2015} mass function.

\citet{Mo1996} developed a method to linearly approximate their CMF by use of a bias function $b_{\rm PS}$. We follow the same steps to calculate a linear bias factor $b_{\rm Trac}$ for the \citet{Trac2015} mass function.
We first substitute equation~(\ref{eq:stdscl}) into equation~(\ref{eq:trac}), and then Taylor expand to linear order about $\delta_b = 0$, and set $\sigma(\massr,z) = 0$. 
We then make a linear volume change correction of $(1+\delta_b)$, giving us
\begin{equation}\label{eq:TLB}
    b_{\rm Trac} = 1+\frac{a}{\delta_{crit}}\frac{(\sigma/2.54)^a}{1+(\sigma/2.54)^a}-\frac{2b}{\sigma^2\delta_{crit}}.
\end{equation}
The CMF can then be approximated as 
\begin{equation}\label{eq:TracLinCMF}
    n_{\rm cond,lin}(\massh,z,\delta_b) = n_{\rm h}(\massh,z)(1+b_{\rm Trac}\delta_b).
\end{equation}
We will use this linear approximation of the CMF to compare to our results when using the full CMF.

For another comparison, we calculate the CMF by scaling the \citet{Trac2015} mass function by the ratio of the conditional to non-conditional Press-Schechter mass functions.
\begin{equation}\label{eq:CMF}
    \frac{n_{\textrm{\rm cond}}}{n_{\rm h}}=\frac{n_{\rm PS,cond}}{n_{\rm PS}}
\end{equation}
where $n_{\rm PS}$ is defined in \citet{Press1974}, and represented here as $f_{\rm PS}(\sigma)$ (which is plugged into eq.~\ref{eq:massfun}).
\begin{equation}
    f_{\rm PS}(\sigma) = \sqrt{\frac{2}{\pi}}\nu~e^{-\nu^2/2}
\end{equation}
The Press-Schechter CMF $n_{\rm PS,cond}(\massh,z,\delta_b, \lagrad)$ is obtained with the ``standard'' coordinate transfer in equation~(\ref{eq:coordtrans}). We then multiply by the same $(1+\delta_r)$ factor to obtain an Eulerian CMF.
We note that this scaling, which was introduced in the high-$z$ context by \citet{Barkana2004}, is commonly used in analytic, semi-analytic, and semi-numeric calculations of galaxy populations at this time (e.g., in \citealt{Mesinger2011}).

At these high redshifts, we ignore the effects of assembly bias (e.g., \citealt{Gao2007}) on the CMF, as it is a small effect compared to the other uncertainties in our model.

%%%%%%%%%%%%%%%%%%%%%%%%%%%%%%%%%%%%%%%%%%%%%%%%%%%%%%%%%%%%%%%%
\subsection{Dark matter density fluctuations}\label{sec:densityfluc}

As stated above, $\sigma(M,z)$ is the linear rms fluctuation of the matter density field at redshift $z$ smoothed over a spherical region of mass $\massr$ (and corresponding $\lagrad$). Thus, the probability distribution of dark matter density for a given scale $\lagrad$ and redshift $z$, $p(\delta_b|\lagrad,z)$, is by definition equal to a zero-mean Gaussian with variance $\sigma^2(\massr,z)$.
However, galaxy surveys measure Eulerian volumes, so to make predictions for them we must convert this distribution to that system. A fixed Eulerian volume will correspond to a range of masses, because each has a different density.
In Appendix \ref{app:EulCorr}, we convert the probability distribution of densities at fixed region mass $p(\delta_b|\lagrad,z)$ to fixed real-space volume $p(\delta_b|\eulrad,z)$. While $p(\delta_b|\lagrad,z)$ is a Gaussian, $p(\delta_b|\eulrad,z)$ is closer to an inverse Gaussian. 
Fortunately, these two distributions are very similar to one another at the region sizes and redshifts we consider in this paper.

However, we do find that $p(\delta_b|\eulrad,z)$ predicts that underdense regions occupy a larger volume-fraction of the Universe than overdense regions at all scales, by as much as $\sim$12 per cent when considering very small scales. This result indicates that surveys will be slightly more likely to probe underdense regions (see Appendix \ref{app:EulCorr} for more details).
Using different methods, \citet{Munoz2010} also found that surveys are more likely to probe an underdense region because of those regions' more rapid cosmic expansion.

\begin{figure}
    \centering
    \includegraphics[width=0.5\textwidth]{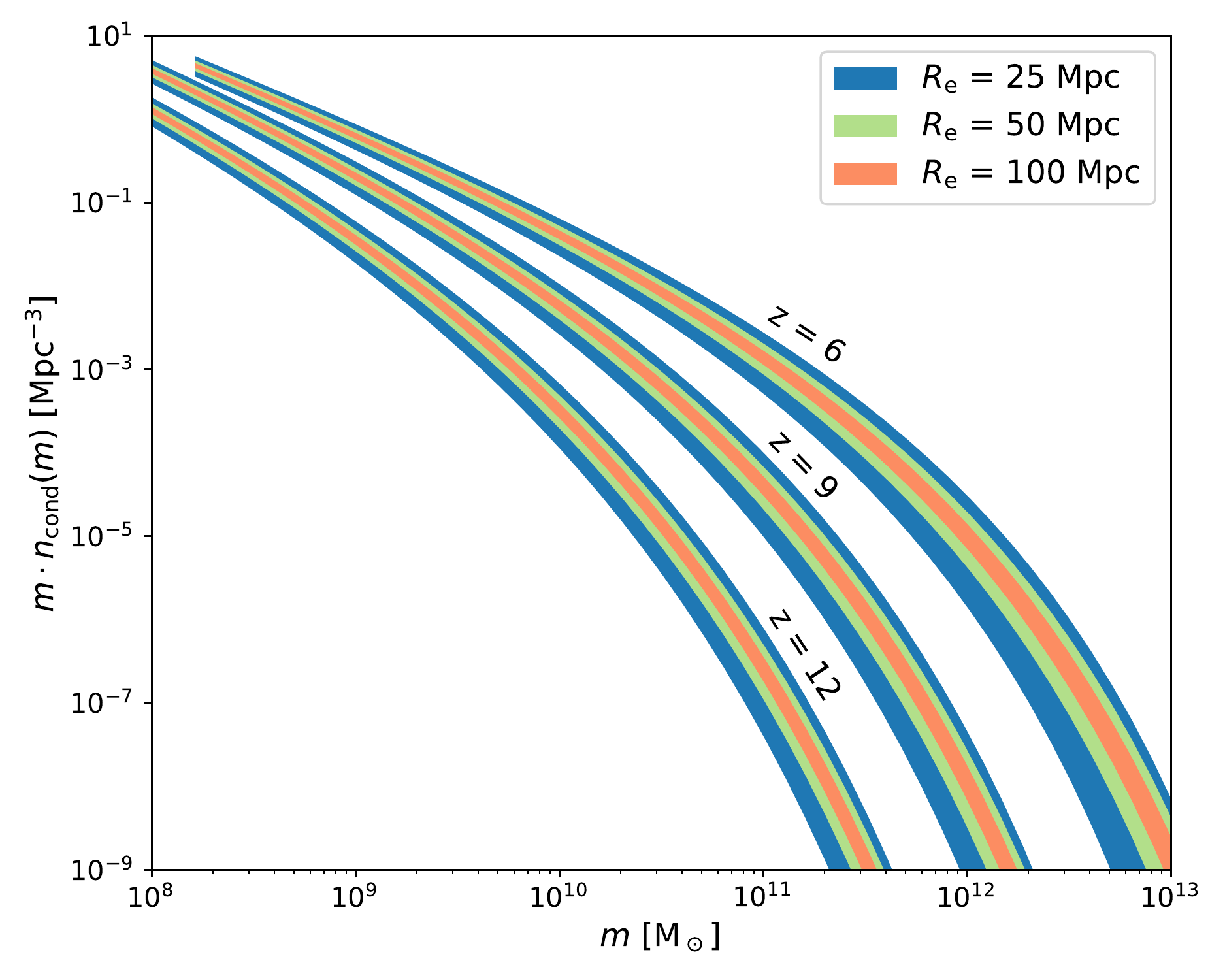}
    \caption{The CMF $n_{\rm cond}$ and its 2$\sigma$ scatter due to cosmic variance on various scales (identified with their radius $\eulrad$) at three redshifts. The scatter increases at high mass and for smaller scales of the Universe (widest shaded area corresponds to the smallest scale).
    }
    \label{fig:CMF}
\end{figure}

With the distribution of densities $p(\delta_b|\eulrad,z)$ and the CMF $n_{\textrm{\rm cond}}(\massh,z,\delta_b, \eulrad)$, we can compute the scatter in halo number density on various scales $\eulrad$ and redshifts.
Figure~\ref{fig:CMF} shows some example results. Cosmic variance in the mass function is substantial for the haloes in which high-$z$ galaxies form. For example, at redshift $z = 9$ on a 50 Mpc (radius) scale, massive haloes ($\sim10^{12}$ M$_\odot$) have a typical relative standard deviation of $\sim$65 per cent, while haloes at the atomic cooling limit (see Section~\ref{sec:feedback}) have a relative standard deviation of $\sim$10 per cent. At fixed halo mass, these relative standard deviations increase at higher redshifts and decrease when considering larger volumes.

%%%%%%%%%%%%%%%%%%%%%%%%%%%%%%%%%%%%%%%%%%%%%%%%%%%%%%%%
\subsection{The CMF in realistic survey volumes}\label{sec:pb}
The CMF presented in section \ref{sec:CMF} assumes a spherical region of radius $\eulrad$. However, real surveys subtend elongated regions pointing away from Earth, commonly referred to as pencil beams. Here, we describe a method for building a CMF for a pencil-beam region. 

We start with the variance in the dark matter density field $\sigma^2$ in a pencil-beam region \citep[following e.g.,][]{Newman2002, Stark2007, Munoz2010, Robertson2010}. For an arbitrarily-shaped volume $\textrm{\textbf{V}}$,
\begin{equation}\label{eq:sigma0fM}
    \sigma^2(\textrm{\textbf{V}}) = \frac{F_g(z)}{(2 \pi)^3} \int P(\textbf{k}) \arrowvert \hat{W}_{\textrm{\textbf{V}}}(\textbf{k}) \arrowvert^2 d\textbf{k},
\end{equation}
where $F_g(z)$ is the growth function (nearly equal to $1/(1+z)$), $\textbf{k}$ is wave vector, $P(\textbf{k})$ is the power spectrum of dark matter \citep[we use the transfer function from][]{Eisenstein1998}, and $\hat{W}_{\textrm{\textbf{V}}}(\textbf{k})$ is the Fourier transform of a real-space top hat in the shape of the region $\textrm{\textbf{V}}$, normalized such that its integral in real-space is equal to unity. In the case of a rectangular pencil-beam volume with side lengths $a_x,~a_y,~a_z$, $\hat{W}_{\textrm{\textbf{V}}}(\textbf{k})=\hat{W}(k_x)\hat{W}(k_y)\hat{W}(k_z)$ with $ \hat{W}(k_i) = \sin(a_i k_i/2)/(a_i k_i/2)$. When constructing a survey volume, we define $a_z$ as the radial distance corresponding to some $\Delta z$ centered at $z$. We define $a_x$ and $a_y$ such that the physical area $a_x*a_y$ at $z$ gives the survey area $A$ as seen from Earth.

We then make the simple approximation that a pencil-beam region has the same CMF as a (larger) spherical region of radius $\regeff$, such that $\sigma_{\rm sphere}(R_{eff}) = \sigma_{\rm PB}(a_x,a_y,a_z)$. This prescription is analogous to how pencil-beam volumes are treated in other analytic studies of cosmic variance. In such studies, $\sigma_{\rm PB}$ is multiplied by a halo bias function to find cosmic variance \citep[e.g.,][]{Stark2007,Munoz2010,Robertson2010,Moster2011}. The halo bias functions used in these studies come from CMFs determined assuming spherical regions, so that pencil-beam volumes are treated as spherical volumes with equivalent $\sigma$.

%%%%%%%%%%%%%%%%%%%%%%%%%%%%%%%%%%%%%%%%%%%%%%%%%%%%%%%%
\subsection{Accretion rates}\label{sec:acc}

We now consider how dark matter haloes accrete matter. This accretion will be used in the next section to determine the rate of star formation.

Many simulations provide similar predictions of halo mass accretion rates \citep[e.g.,][]{McBride2009, Fakhouri2010, vandenBosch2014, Goerdt2015, Trac2015}. However, these rates have not been tested at the very high redshifts and very low masses relevant to our model. For this paper, we calculate the accretion rates using the method described in \citet{Furlanetto2017}, which is analogous to abundance matching \citep{Vale2004}: haloes maintain a constant number density as they evolve according to the mass function of the region they are in.
That is, we require that at any two nearby redshifts $z_1$ and $z_2$, a halo has masses $\massh_1(z_1)$ and $\massh_2(z_2)$ such that:
\begin{equation}\label{eq:acc}
\begin{aligned}
    &\int_{\massh_1}^{\infty}d\massh~n_{\textrm{\rm cond}}(\massh,z_1,\delta_b,\lagrad)~=\\
    &\int_{\massh_2}^{\infty}d\massh~n_{\textrm{\rm cond}}(\massh,z_2,\delta_b,\lagrad),
\end{aligned}
\end{equation}
where $n_{\textrm{\rm cond}}(\massh,z,\delta_b,\lagrad)$ is the Lagrangian CMF from equation (\ref{eq:CMF}). We define the accretion rate of haloes $\dot{\massh}_{\rm h}$ in a region such that they satisfy equation~(\ref{eq:acc}) at all masses over a small redshift interval ($\Delta z \sim 0.1$).

In practice this accretion method means that in a given region, the most massive halo at one time step is also the most massive halo at the next time step, and the same goes for the second and third most massive haloes, etc.
In this treatment, accretion is continuous and smooth, increases monotonically with halo mass, and has zero scatter at a fixed mass.
This treatment is obviously not entirely correct, but it is in line with our goal of simplicity and has the added benefit maintaining the CMF across cosmic time in a way that conserves mass. \citet{Furlanetto2017} show that accretion rates obtained using this method are similar to the simulation accretion rates mentioned above in the redshift and mass ranges they probe. We do neglect mergers in this model, which will provide an additional source of scatter (see the discussion in \citealt{Furlanetto2017}).

\begin{figure}
    \centering
    \includegraphics[width=0.5\textwidth]{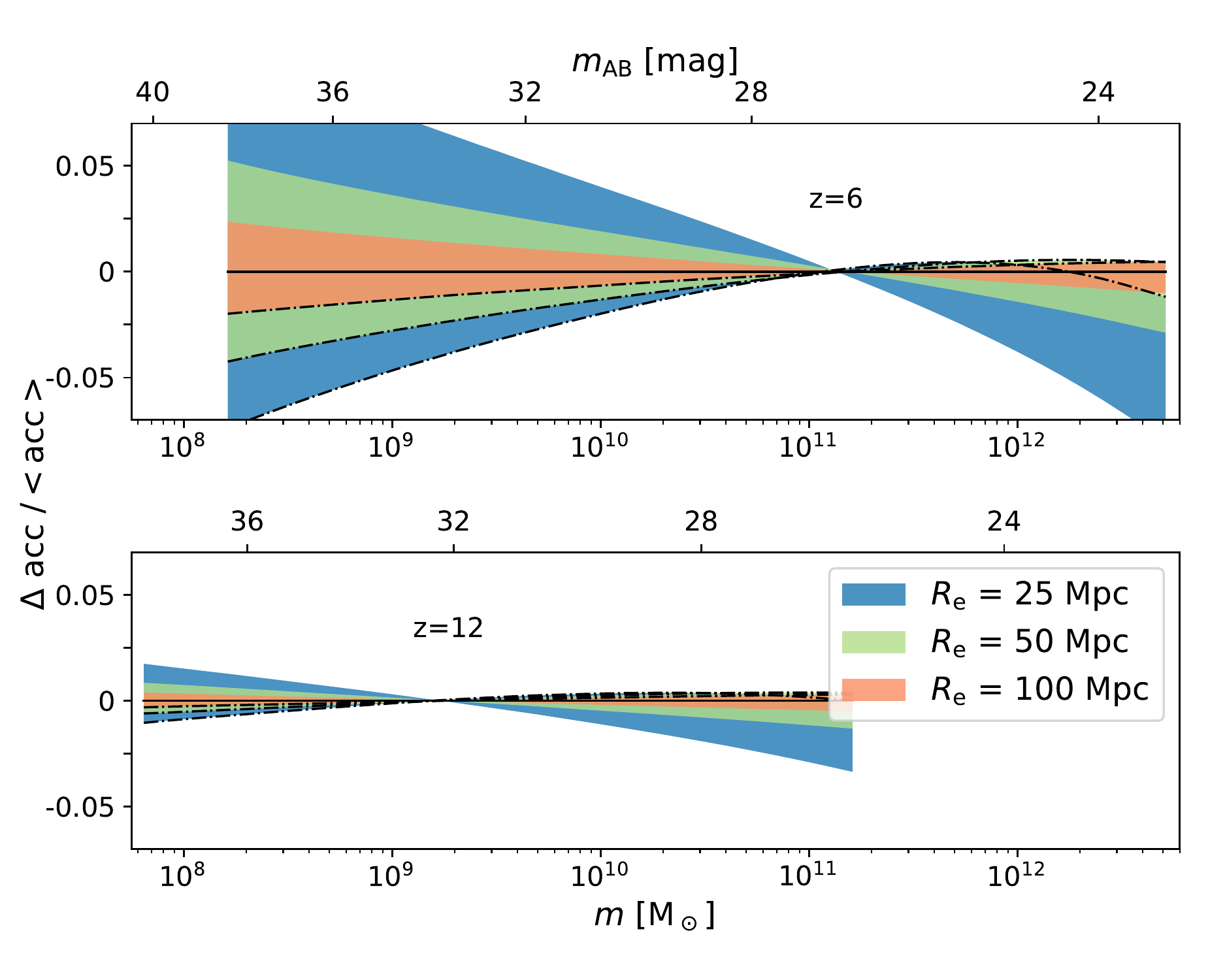}
    \caption{Accretion rate scatter (shaded areas) due to cosmic variance compared to the average accretion rate of haloes (solid black line) for various scales (identified with their radius $\eulrad$). Adding cosmic variance to our models changes accretion rates by $<5$ per cent for most haloes; smaller regions have larger variance in accretion. The dot dashed lines show the accretion of a 1$\sigma$ overdense region, and the opposite end of the shaded regions show the 1$\sigma$ underdense region. Higher mass haloes overaccrete in overdense environments and underaccrete in underdense environments. For lower mass haloes, the opposite is true. The threshold mass where haloes in all environments accrete nearly equally evolves to lower mass with redshift. The top axes show the approximate apparent magnitude of the haloes \textquotedblleft$\mapp$" (we assign haloes their magnitudes in section \ref{sec:FRSF}).
    }
    \label{fig:acc}
\end{figure}

This method allows haloes in over and underdense regions to accrete at different rates. However, we find that this is actually a small effect. The shaded areas in Figure~\ref{fig:acc} show how the accretion rates depend on large scale environment. For most masses, redshifts, and scales, the variation in accretion is less than 5 per cent. A 5 per cent difference in accretion can be significant over a Hubble time, but we will show it has a small effect on the UVLF, which is most sensitive to instantaneous star formation.

%%%%%%%%%%%%%%%%%%%%%%%%%%%%%%%%%%%%%%%%%%%%%%%%%%%%%%%%%%%%%%%%
%%%%%%%%%%%%%%%%%%%%%%%%%%%%%%%%%%%%%%%%%%%%%%%%%%%%%%%%%%%%%%%%

%%%%%%%%%%%%%%%%%%%%%%%%%%%%%%%%%%%%%%%%%%%%%%%%%%%%%%%%%%%%%%%%
%%%%%%%%%%%%%%%%%%%%%%%%%%%%%%%%%%%%%%%%%%%%%%%%%%%%%%%%%%%%%%%%
\section{Feedback-Regulated Star Formation}\label{sec:FRSF}

In this section we transform the mass accretion rates of haloes into ultraviolet (UV) luminosities. \citet{Furlanetto2017} provides a detailed explanation of our star formation model; we briefly summarize it here. We intentionally choose this simple, ``minimalist" model so as to make our assumptions about the mass-luminosity relation transparent. Estimates of cosmic variance must necessarily account for the many uncertainties about high-$z$ galaxies, and a simple, flexible model allows us to estimate how important the specifics of galaxy formation are for the variance.

%%%%%%%%%%%%%%%%%%%%%%%%%%%%%%%%%%%%%%%%%%%%%%%%%%%%%%%%
\subsection{Models of feedback}\label{sec:feedback}

We assume that haloes only form stars when they exceed a threshold  mass $\massh_{\rm min}$. This mass corresponds to a halo virial temperature $T_{vir}=10^4K$, when atomic line cooling becomes efficient enough for gas clouds to collapse and fragment for star formation \citep{Loeb2013}. This mass is typically $\massh_{\rm min}\sim10^8M_{\odot}$. At the redshifts considered in this paper, haloes at the threshold are always far below the detection limit of next generation telescopes.

Gas accreting onto a galaxy can be turned into stars. When stars form, they expel baryons from their host galaxy through radiation pressure, supernovae, or some other process like grain heating \citep[e.g., ][]{FaucherGiguere2013, Hayward2017, Krumholz2018}. Balancing this stellar feedback with accretion provides a simple estimate of the star formation rate $\dot{\massh}_*$ of a galaxy via
\begin{equation}
    \dot{\massh}_* =~\dot{\massh}_{\rm b} - \dot{\massh}_{\rm w},
\end{equation}
where $\dot{\massh}_{\rm b}$ is the mass accretion rate of the halo times the baryon fraction $\dot{\massh}_{\rm b} = [\Omega_{\rm b}/\Omega_{\rm m}]\dot{\massh}_{\rm h}$, and $\dot{\massh}_{\rm w}$ is the rate of baryon loss through feedback. The fraction of accreting baryons that are converted into stars is defined as $f_* = \dot{\massh}_*/\dot{\massh}_{\rm b}$. Finally, we write the mass ejection rate as a multiple of the star formation rate $\dot{\massh}_{\rm w} = \eta(\massh,z)~\dot{\massh}_*$, yielding
\begin{equation}\label{eq:fstarONE}
    f_*=\frac{1}{1+\eta(\massh,z)}.
\end{equation}

Many models suggest that massive haloes accrete gas more slowly than our simple argument suggests, because of the heating at the virial shock.
\citet{Furlanetto2017} show that virial shock heating only has a modest effect on the results of this model, but we include it because it helps match the observed densities at large luminosities. \citet{FaucherGiguere2011} show the fraction of gas that can cool onto a galaxy in the presence of a virial shock is \footnote{We require $f_{shock} \leq 1$, and we smooth the function near where $f_{shock}\rightarrow1$ in order to ensure that it is smoothly differentiable.}
\begin{equation}
    f_{shock}=0.47\left(\frac{1+z}{4}\right)^{0.38}\left(\frac{\massh}{10^{12}M_\odot}\right)^{-0.25}.
\end{equation}

Because we only include stellar feedback, which limits star formation at small masses, we 
also impose a maximum efficiency $f_{*,max}$ that limits star formation when $\eta(\massh,z)\rightarrow0$ at large halo masses. We impose it in a way that keeps $f_*$ smoothly differentiable.
Combining $f_{shock}$ and $f_{*,max}$ with equation \ref{eq:fstarONE} gives
\begin{equation}
    f_*=\frac{f_{shock}}{f_{*,max}^{-1}+\eta(\massh,z)}.
\end{equation}

Finally, we parameterize the strength of stellar feedback, $\eta(\massh,z)$, as
\begin{equation}
    \eta=C\left(\frac{10^{11.5}M_\odot}{\massh}\right)^\xi \left(\frac{9}{1+z}\right)^\sigma.
\end{equation}
For energy driven supernova feedback: $C=1$, $\xi=2/3$, $\sigma=1$, and $f_{*,max}=0.1$. We will also consider a redshift-independent version ($C=2$, $\xi=2/3$, $\sigma=0$, and $f_{*,max}=0.1$) and a momentum-driven version ($C=5$, $\xi=1/3$, $\sigma=1/2$, and $f_{*,max}=0.2$) for comparison \citep[for more details on $\eta$ and its parameterization, see][]{Sun2016,Mirocha2017,Furlanetto2017}. These alternate parameterizations of $\eta$ will allow us to test how cosmic variance depends on our galaxy formation model.

This simple model undoubtedly ignores many important elements of galaxy formation, but it suffices to consider a wide range of possible halo mass-luminosity relations.
For example, we do not take into account that gas should cycle through the interstellar medium (ISM) before forming stars. However, in the `bath tub' model of galaxy formation, galaxies evolve towards a quasi-equilibrium state between mass accretion and star formation such that the ISM maintains roughly constant mass \citep{Dekel2014}. Once this equilibrium is reached, our model more accurately describes star formation.

%%%%%%%%%%%%%%%%%%%%%%%%%%%%%%%%%%%%%%%%%%%%%%%%%%%%%%%%
\subsection{From star formation to luminosity}

We now convert star formation rate to UV luminosity. UV luminosity is a good tracer of star formation because it is produced only by massive, short-lived stars. We take the standard conversion
\begin{equation}\label{eq:SFconv}
    \dot{\massh}_* = K_{UV} \times L_{UV},
\end{equation}
where $L_{UV}$\footnote{For the remainder of the paper, we will display luminosity as absolute and apparent AB magnitudes ($\mabs$ and $\mapp$).} is the rest-frame continuum ($1500-2800$ \AA)\footnote{This wavelength range corresponds to 
$H$-band in the redshift range of $z\approx5-9$, and $K$-band for $z\approx8-12$.} intrinsic luminosity (without extinction). $K_{UV}$ is a conversion from luminosity to star formation rate, and it is dependent on the initial mass function, metallicity, star formation history, binaries, etc. We take $K_{UV}=1.15\times10^{-28} M_\odot yr^{-1}/(erg~s^{-1} Hz^{-1})$ from \citet{Madau2014}. We will show that $K_{UV}$ will not substantially affect our predictions for the relative cosmic variance of the UVLF, even though it can have significant effects on the UVLF itself.

We do ignore dust in our fiducial model, because the extinction in these sources is only poorly constrained. Models suggest that it is modest, and most importantly is not a strong function of halo mass \citep{Mirocha2020}. We do however test the effects of dust on our results at $z < 8$ using an empirical dust correction \citep[][``Model A'']{Vogelsberger2020}. In order to match the data when applying this dust correction to our energy-driven model, we set $C = 2$ and $f_{*,max} = 0.3$, making the galaxies intrinsically brighter at fixed halo mass. For simplicity, we ignore $f_{shock}$ in this case, because it also affects the bright end of the luminosity function.

We also ignore scatter in the halo mass to luminosity relation, which would have the effect of flattening out the exponential drop off of the UVLF, as upward scatter in the luminosity has a larger relative effect on the luminosity function in that regime. While we expect this effect to be small (at least on population-level statistics such as the UVLF), we plan to explore it in the future by introducing a scatter in the accretion rates and/or star formation rates.

%%%%%%%%%%%%%%%%%%%%%%%%%%%%%%%%%%%%%%%%%%%%%%%%%%%%%%%%%%%%%%%%
%%%%%%%%%%%%%%%%%%%%%%%%%%%%%%%%%%%%%%%%%%%%%%%%%%%%%%%%%%%%%%%%

%%%%%%%%%%%%%%%%%%%%%%%%%%%%%%%%%%%%%%%%%%%%%%%%%%%%%%%%%%%%%%%%
%%%%%%%%%%%%%%%%%%%%%%%%%%%%%%%%%%%%%%%%%%%%%%%%%%%%%%%%%%%%%%%%
\section{Cosmic Variance in the UVLF} \label{sec:cosvar}

In this section we present the conditional UVLF generated by our model. We provide a fit to the conditional UVLF with a simple Gaussian approximation. We then test the robustness of our results against model choices. Finally, we compare our results to recent works.

\begin{figure}
    \centering
    \includegraphics[width=0.5\textwidth]{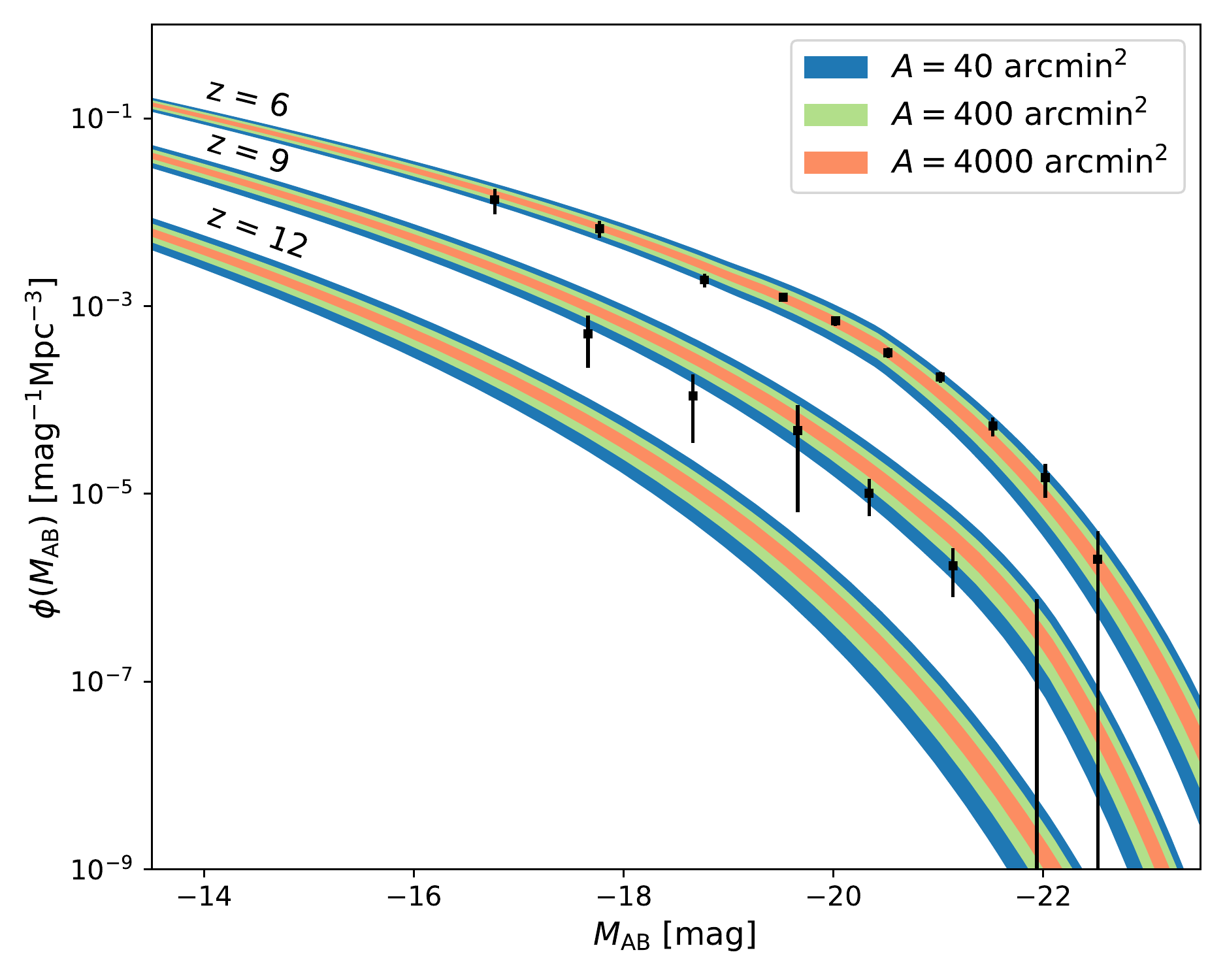}
    \caption{The UVLF and its 2$\sigma$ scatter from cosmic variance for three selections of survey area at three redshifts ($z$ = 6, 9, and 12; $z$ = 9 and 12 are off-set in log space by -0.5 dex and -1 dex, respectively). The scatter in the UVLF increases at the bright end and for smaller survey areas, similar to the CMF in Figure~\ref{fig:CMF}. The survey areas $A = 40,$ $400,$ $4000$ arcmin$^2$ have volumes equivalent to spheres with radii $\eulrad \approx 29,$ $63,$ $135$ Mpc at $z=6$ (redshift bin $\Delta z = 1$). The data points are from \citet{Bouwens2015} and \citet{Bouwens2016}.}
    \label{fig:CLF}
\end{figure}

We show the conditional UVLF $\phi_{\rm cond}(\mabs,z,\delta_b,\eulrad)$ and its 2$\sigma$ scatter due to cosmic variance in Figure~\ref{fig:CLF}. As in the CMF, cosmic variance increases with increasing galaxy luminosity and also with increasing redshift. The data points shown in Figure~\ref{fig:CLF} are from \citet{Bouwens2015} and \citet{Bouwens2016}; for a more in-depth analysis of this model's agreement with current data, see \citet{Furlanetto2017}.

The mapping from halo mass to luminosity in our model is nearly independent of environment because accretion is also nearly independent of environment (see Fig.~\ref{fig:acc}). Thus, nearly all of the cosmic variance of the UVLF comes directly from the variance in the CMF (see Fig.~\ref{fig:CMF}). 
%AT: Adding Lovell
Similarly, simulations by \citet{Lovell2020} find the star formation rate of a galaxy is independent of the dark matter environment, although, they compare star formation rates at fixed stellar mass, not halo mass.

%%%%%%%%%%%%%%%%%%%%%%%%%%%%%%%%%%%%%%%%%%%%%%%%%%%%%%%%

\subsection{Calculating cosmic variance}\label{sec:powerfit}

\begin{figure}
    \centering
    \includegraphics[width=0.5\textwidth]{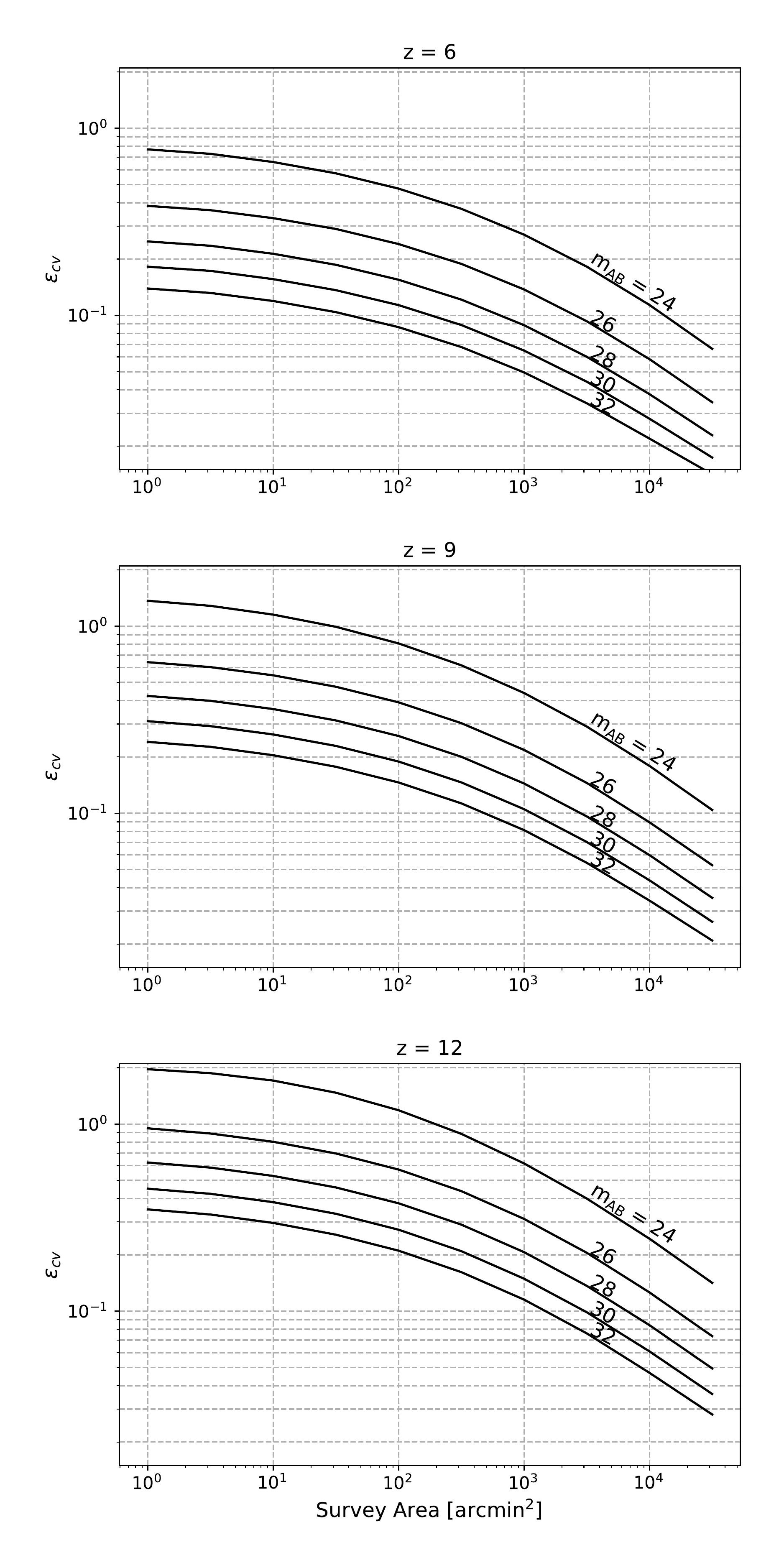}
    \caption{The relative cosmic variance of the UVLF $\varepsilon_{cv}$ as a function of survey area for various apparent magnitudes with a redshift bin width of $\Delta z = 1$ (\textit{black lines}; shown from apparent magnitude $m_{\rm AB} = 32$ on the bottom, decreasing by $\Delta m_{\rm AB} = 2$ towards the top).}
    \label{fig:powerfit}
\end{figure}

As shown in Figure~\ref{fig:CLF}, the amount of cosmic variance in a given galaxy formation model will depend on luminosity, redshift, and the survey characteristics. In this section we provide a simple descriptor of cosmic variance across all these parameters. We quantify cosmic variance $\varepsilon_{cv}$ as the relative standard deviation of the conditional UVLF at fixed redshift, apparent magnitude, survey area, and redshift bin width\footnote{Our definition of $\mapp$ assumes the galaxy is at the specified `fixed redshift', regardless of redshift bin width.}:
\begin{equation}
    \varepsilon^2_{cv}=\frac{\langle \phi_{\rm cond}^2 \rangle - \langle \phi_{\rm cond} \rangle^2}{\langle \phi_{\rm cond} \rangle^2},
\end{equation}
with $\langle \phi_{\rm cond}^n \rangle$ defined as
\begin{equation}
    \langle \phi_{\rm cond}^n \rangle = \int \phi_{\rm cond}^n(\mapp,z,\delta_b,\eulrad) \times p(\delta_b|\eulrad,z) d\delta_b,
\end{equation}
where $\eulrad$ is determined from the survey area and redshift bin width as described in Section~\ref{sec:pb}. Figure~\ref{fig:powerfit} shows $\varepsilon_{cv}$ as a function of survey area for various redshifts and apparent magnitudes (all with $\Delta z = 1$). This definition of $\varepsilon_{cv}$ uses $\phi_{\rm cond}$ at fixed $z$, but applies it to the entire volume defined by $A$ and $\Delta z$. This approximation breaks down if cosmic variance evolves significantly over the range defined by $\Delta z$. Thus, the choice of $\Delta z$ should be made with care, especially at lower $z$ where $\varepsilon_{cv}$ evolves most rapidly in a relative sense ($\varepsilon_{cv}$ evolves more rapidly in an absolute sense at high $z$). Over $\Delta z = 1$, $\varepsilon_{cv}$ evolves 10\% -- 30\% (at $z$ = 14 \& 5, respectively). Choosing $\Delta z < 0.5$ keeps the change in $\varepsilon_{cv}$ below 10\% for most cases\footnote{Our public Python package \pakidge~can be used to explore the evolution of $\varepsilon_{cv}$ over any desired parameter.}.

\begin{table}
	\centering
	\caption{Parameters for fits to $\varepsilon_{cv}$ (eq.~\ref{eq:powerfit}). We provide $\varepsilon_{cv}$ on a much wider range of parameters via our public Python package \pakidge~(see Data Availability section).}
	\label{tab:powerfit}
	\begin{tabular}{ccccc}
		\hline
		\hline
		Redshift & App. UV mag. &  & Fit parameters & \\
		$z$ & m$_{\rm AB}$ & $\Psi$ & $\gamma$ & $b$\\
		\hline
        6 & 32 & -0.223 & 0.167 & -0.608 \\
          & 30 & -0.189 & 0.184 & -0.529 \\
          & 28 & -0.174 & 0.192 & -0.410 \\
          & 26 & -0.161 & 0.200 & -0.235\\
          & 24 & -0.165 & 0.199 & 0.070\\
        \hline
        9 & 32 & -0.198 & 0.184 & -0.399\\
          & 30 & -0.190 & 0.188 & -0.298\\
          & 28 & -0.178 & 0.194 & -0.175\\
          & 26 & -0.173 & 0.197 & -0.002\\
          & 24 & -0.197 & 0.188 & 0.353\\
        \hline
        12 & 32 & -0.195 & 0.188 & -0.240\\
           & 30 & -0.188 & 0.191 & -0.137\\
           & 28 & -0.184 & 0.193 & -0.001\\
           & 26 & -0.185 & 0.193 & 0.182\\ 
           & 24 & -0.202 & 0.189 & 0.533\\
        \hline
	\end{tabular}
\end{table}

Figure~\ref{fig:powerfit} shows that the relative importance of cosmic variance varies widely across the galaxy population, with a strong dependence on survey parameters. Using a redshift bin width $\Delta z = 1$, $\varepsilon_{cv}$ is low at the faint end of the UVLF ($m_{\rm AB} = 32$), ranging from $\sim5$\% at large survey area (1000 arcmin$^2$) to $\sim15$\% at small survey area (1 arcmin$^2$) at $z=6$. As redshift increases, so does $\varepsilon_{cv}$; at $z=12$ and $m_{\rm AB} = 32$, $\varepsilon_{cv}$ ranges from $\sim12$\% at large survey area (1000 arcmin$^2$) to $\sim35$\% at small survey area (1 arcmin$^2$). 

Cosmic variance also increases significantly at the bright end of the UVLF. At $m_{\rm AB} = 26$, $\varepsilon_{cv}$ ranges from $\sim15$\% at large survey area (1000 arcmin$^2$) to $\sim40$\% at small survey area (1 arcmin$^2$) at $z=6$. At $z=12$ and $m_{\rm AB} = 26$, $\varepsilon_{cv}$ ranges from $\sim30$\% at large survey area (1000 arcmin$^2$) to $\sim90$\% at small survey area (1 arcmin$^2$).

Cosmic variance flattens out at small survey areas, which is largely due to the effects of the pencil-beam shape of surveys.
Even at small survey areas, such a geometry still contains a relatively large range of environments due to its elongated shape. This effect keeps cosmic variance much lower than what one would obtain with a spherical region of the same volume.

We approximate $\varepsilon_{cv}$ with a simple functional form; a polynomial in log$_{10}(\varepsilon_{cv})$ fits well:
\begin{equation}\label{eq:powerfit}
    \textrm{log}_{10}(\varepsilon_{cv}) \approx \Psi A^\gamma+b,
\end{equation}
where $A$ is in arcmin$^2$, and $\Psi,~\gamma,\textrm{and }b$ are fit parameters\footnote{Note that our fit assumes the survey subtends a square area on the sky.}.
Table \ref{tab:powerfit} displays the parameter fits at a selection of redshifts and magnitudes. These fits have a typical/maximum fractional error of 3/5\%. We provide $\varepsilon_{cv}$ for a wider range of parameters via a public python package \pakidge~(see Data Availability section for more details).

With $\varepsilon_{cv}$, we define a linear approximation of the conditional UVLF in a region with density $\delta_b$, angular extent $A$, and redshift bin width $\Delta z$:
\begin{equation}\label{eq:CLF}
\begin{aligned}
    &\phi_{\rm cond}(\mapp,z,\delta_b,A,\Delta z)=\\
    &\langle \phi(\mapp,z) \rangle\left[1+\varepsilon_{cv}(A,\mapp,z,\Delta z)\frac{\delta_b}{\sigma_{\rm PB}}\right],
\end{aligned}
\end{equation}
where $\langle \phi(\mapp,z) \rangle$ is the average UVLF and $\delta_b/\sigma_{\rm PB}$ is the density of the region relative to a 1$\sigma$ fluctuation. This conditional UVLF is similar in construction to that in \citet{Livermore2017}, used to fit to lensed high-z galaxy survey data.

%%%%%%%%%%%%%%%%%%%%%%%%%%%%%%%%%%%%%%%%%%%%%%%%%%%%%%%%
\subsection{Parameter dependence of $\varepsilon_{cv}$}\label{sec:uncertainty}

Our calculations so far have assumed our fiducial choices for the galaxy model (assuming the minimalist energy-regulated prescription) and mass function parameters. Here, we explore how sensitive our results are to variations in these assumptions.

\begin{figure}
    \centering
    \includegraphics[width=0.5\textwidth]{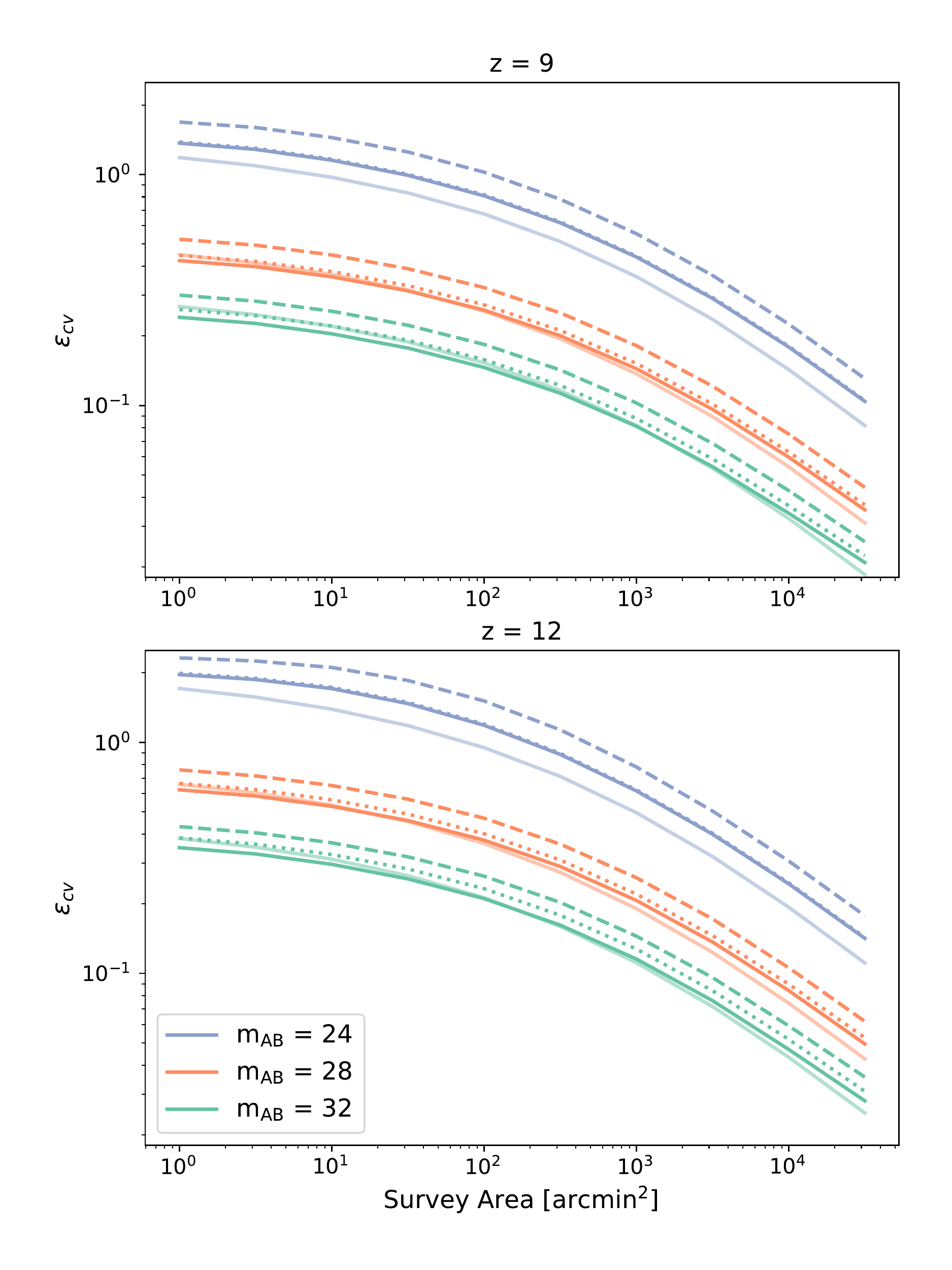}
    \caption{The relative cosmic variance of the UVLF $\varepsilon_{cv}$ as a function of survey area for various apparent magnitudes (\textit{solid lines}, same as Fig.~\ref{fig:powerfit}). The \textit{dotted lines} show the effects of switching to a z-independent version of star formation. The \textit{faded solid lines} show the linear bias method for estimating cosmic variance (eq.~\ref{eq:epCVlin}).  The \textit{dashed lines} show the effects of using a different method for creating the CMF, specifically the ``Press-Schechter scaling'' approach applied to the \citet{Trac2015} mass function (see eq.~\ref{eq:CMF}). The three sets of lines correspond to magnitudes 32, 28, and 24 (bottom to top).}
    \label{fig:epCVcompare}
\end{figure}

First, we consider how cosmic variance depends on the star formation model. Figure~\ref{fig:epCVcompare} shows shows the difference in $\varepsilon_{cv}$ when using our fiducial energy-regulated feedback (solid lines) vs a redshift-independent version of feedback (dotted lines, see Section~\ref{sec:feedback}). There is little difference in the predictions for $\varepsilon_{cv}$. We find a similarly small difference when using momentum-regulated feedback and when using the dust correction from \citet{Vogelsberger2020} for $z < 8$ (the effects of these two are similar to the dotted lines; so are not plotted to reduce clutter). Also, our choice of $K_{UV}$ (see eq.~\ref{eq:SFconv}) will not significantly affect our results, as $\varepsilon_{cv}$ is not a particularly strong function of magnitude.
These results suggest that cosmic variance is not strongly dependent on the details of star formation or dust.

Second, we explore if cosmic variance is strongly affected by large-scale galaxy environment, namely through differences in accretion. With the linear halo bias function from equation~(\ref{eq:TLB}), we approximate 
\begin{equation}\label{eq:epCVlin}
    \varepsilon_{cv} \approx b_{\rm Trac}\sigma_{\rm PB},
\end{equation}
and show it in Figure~\ref{fig:epCVcompare} (faded solid line)\footnote{We connect the bias function $b_{\rm Trac}$ to galaxies using our model's average halo mass--UV luminosity relation}. While our full model allows for galaxies to have an environment-dependent accretion and thus luminosity, this linear method does not. However, it provides very similar results to the full method, though it slightly underpredicts cosmic variance at the bright end and overpredicts at the faint end due to the variance in accretion for those haloes (see Fig.~\ref{fig:acc}). This result suggests that approximating the CMF via a simple bias factor is sufficient to capture the effects of cosmic variance (at least to linear order; large density excursions are discussed later in this section).

We conclude that the level of cosmic variance is not sensitive to the particulars of the galaxy formation model. Rather, cosmic variance is dominated by the underlying conditional halo mass function. In section~\ref{sec:CMF}, we described an alternate method of creating a CMF: scaling the \citet{Trac2015} mass function by the conditional \citet{Press1974} mass function (eq.~\ref{eq:CMF}). In Figure~\ref{fig:epCVcompare} we show $\varepsilon_{cv}$ when using that CMF (dashed lines). This change results in $\sim25$\% more cosmic variance across the board, the largest effect of any model choice. Thus, in our model, the biggest uncertainty in $\varepsilon_{cv}$ is in our understanding of the CMF.

\begin{figure*}
    \centering
    \includegraphics[width=7.0in]{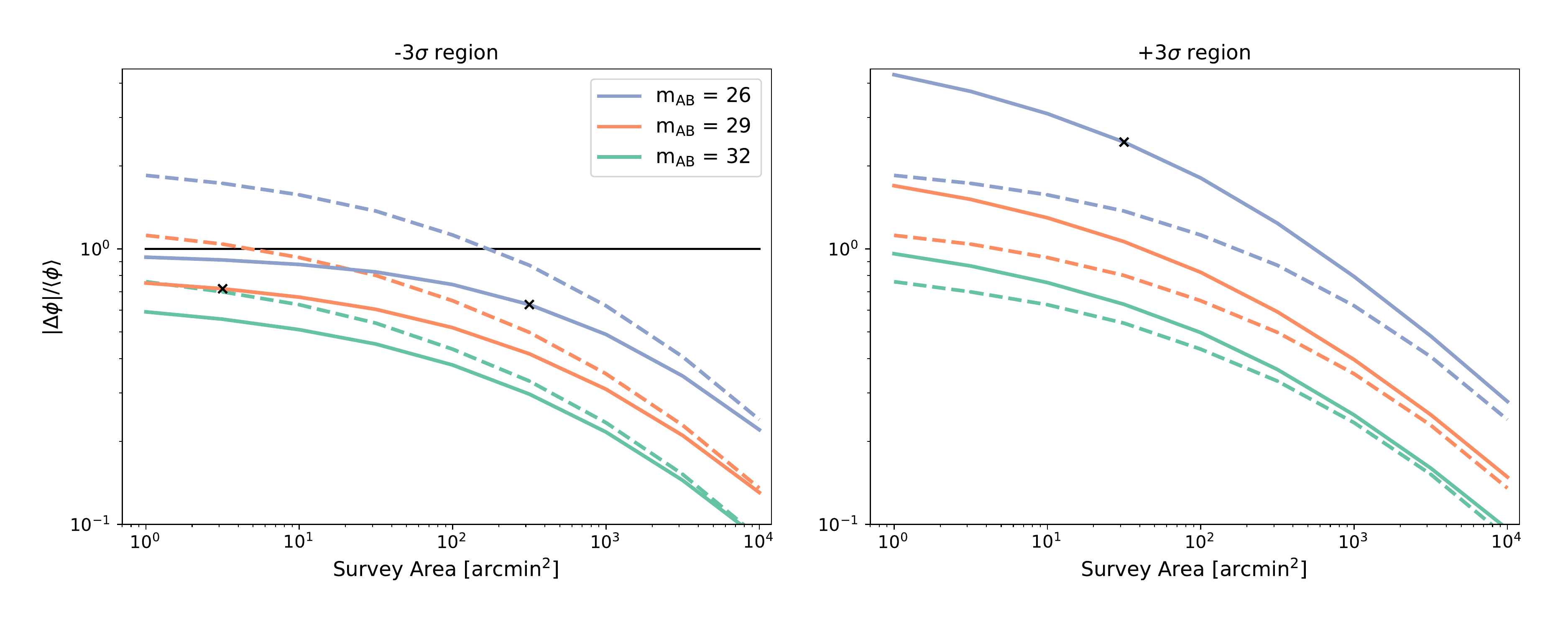}
    \caption{The relative excursion from the average UVLF $\langle \phi \rangle$ for a 3$\sigma$ underdense region (\textit{left}) and a 3$\sigma$ overdense region (\textit{right}) at redshift 9. We compare the full treatment (\textit{solid lines}) and the linear approximation in equation~(\ref{eq:CLF}) (\textit{dashed lines}). In the left panel, the linear approximation predicts there will be fewer galaxies than the full approach, and can even predict unphysical negative galaxy number densities (when above the horizontal black line). In the right panel, again the linear approximation underpredicts the expected number of galaxies. The black 'x' marks the survey area where our model predicts there to be $\sim$1 source in the corresponding magnitude bin. The solid and dashed lines show apparent magnitudes 32, 29, and 26 (bottom to top).}
    \label{fig:epCV_3Sig}
\end{figure*}

While equation~(\ref{eq:CLF}) provides a good approximation to the conditional UVLF for $\delta_b/\sigma_{\rm PB} \lesssim 2$, the assumption of a Gaussian bias distribution breaks down at larger density excursions. Figure~\ref{fig:epCV_3Sig} shows (at $z = 9$) the difference between using equation~(\ref{eq:CLF}) (\textit{dashed lines)} and our full treatment (\textit{solid lines}) for the conditional UVLF ($\phi$) for a 3$\sigma$ density excursion. Equation~(\ref{eq:CLF}) underestimates the number of galaxies in very underdense regions (even giving unphysical negative densities at the smallest survey areas), and it also underestimates the number of galaxies in very overdense regions. Equation~(\ref{eq:CLF}) underpredicts the number of galaxies in both wings, even though it is more reliable near $\delta_b = 0$,  because the true bias distribution (at fixed magnitude) is closer to a log-normal. However, where the deviation from the Gaussian approximation is most pronounced, Poisson shot noise usually dominates the error. Thus, for most applications, equation~(\ref{eq:CLF}) ($\varepsilon_{cv}$) adequately captures the behaviour of the conditional UVLF.

%%%%%%%%%%%%%%%%%%%%%%%%%%%%%%%%%%%%%%%%%%%%%%%%%%%%%%%%
\subsection{Comparison to other works}

Here we compare our predictions of cosmic variance to those from two recent models in the literature: \citet{Bhowmick2020} and \citet{Ucci2020}.

\citet{Bhowmick2020} provide public estimates for cosmic variance in a redshift range $z = 7$--14 and for apparent $H$-band magnitudes between $\mapp= 25$ and 30. They determine cosmic variance first by calculating the two-point correlation function of galaxies in their simulation box. They then fit the correlation function to a power law and integrate it in a pencil-beam volume (see their eq. 2) to estimate the relative cosmic variance .
They provide estimates of cosmic variance for all sources \textit{brighter} than the listed magnitude, rather than for sources \textit{at} the listed magnitude. This choice means their estimates of cosmic variance are higher than they would be at fixed magnitude, as cosmic variance increases for brighter sources. However, when we mimic this cumulative method, we find the effect is relatively small (cosmic variance $\lesssim10$\% larger than fixed magnitude method).

\citet{Ucci2020} provide public estimates for cosmic variance in a redshift range $z=6$--12 and for apparent magnitudes $\mapp=24$--38 (at $z = 9$). 
\citet{Ucci2020} calculate cosmic variance as the relative standard deviation of galaxy number counts in many pencil-beam sub-volumes of their simulation box.
They have a slightly more limited survey area coverage, providing estimates between $A=1$ and 1000 arcmin$^2$. We compare to the predictions from their ``photoionization'' model. These predictions include Poisson variance, making them an estimate of the total variance rather than just cosmic variance.

\begin{figure}
    \centering
    \includegraphics[width=0.5\textwidth]{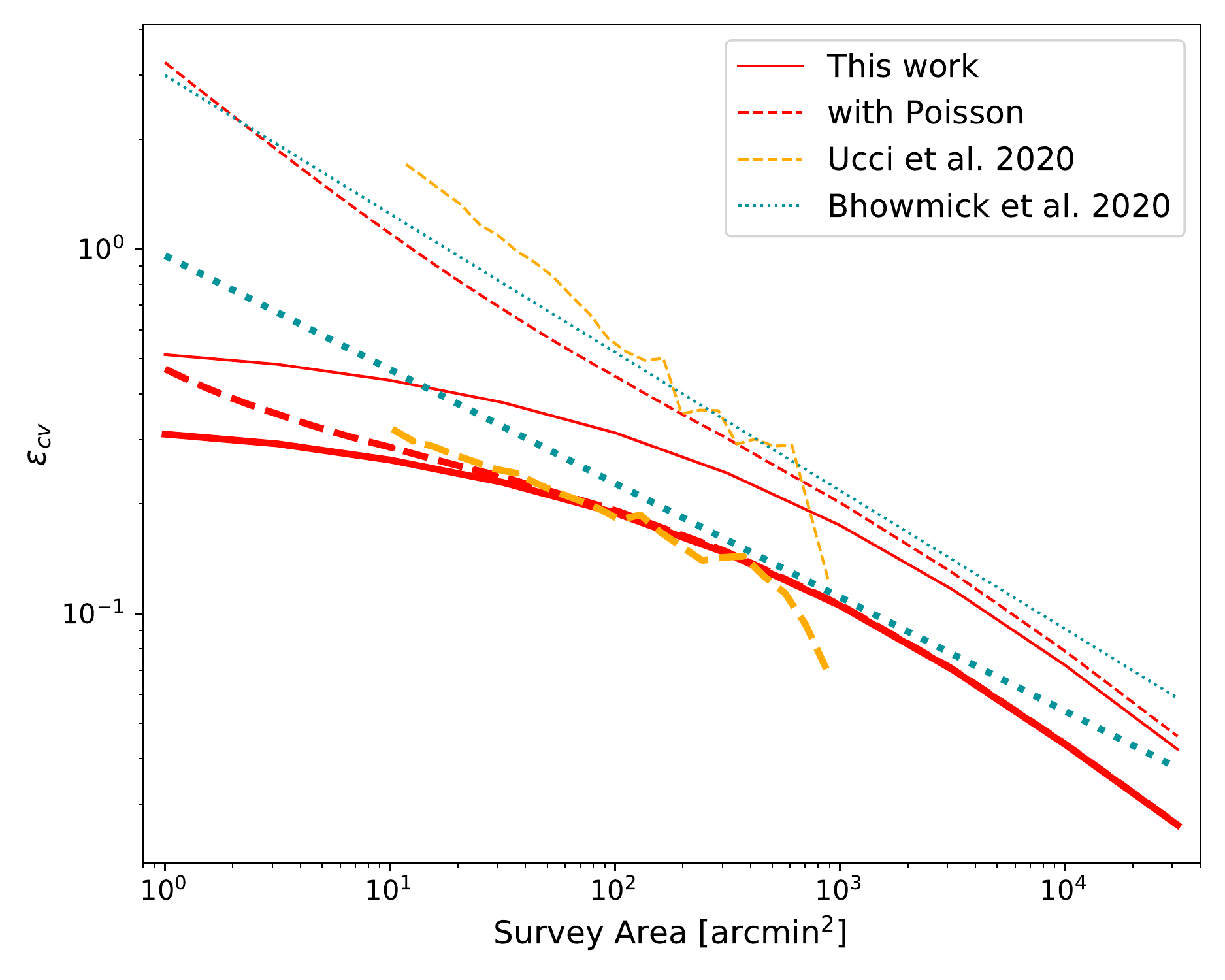}
    \caption{Comparison of cosmic variance predictions at $z$ = 9 (with a redshift window of $\Delta z = 1$). The lower set of lines (thick) is at an apparent magnitude of $\mapp = 30$, while the upper set of lines (thin) is at $\mapp = 27$. The dashed red lines show our cosmic variance predictions with Poisson noise added, for comparison with \citet{Ucci2020}. 
    }
    \label{fig:litComp_z9}
\end{figure}

Figure~\ref{fig:litComp_z9} shows our predictions compared to those of \citet{Bhowmick2020} (blue dotted lines) and \citet{Ucci2020} (yellow dashed lines) at $z = 9$, with a redshift window of $\Delta z = 1$, at apparent magnitudes of $\mapp = 30$ (lower, thick curves) and $\mapp=27$ (upper, thin curves). 

As \citet{Ucci2020} report total variance, their predictions should be compared with the red dashed lines (our prediction plus Poisson noise). Our predictions agree closely with those of \citet{Ucci2020} at $\mapp=30$, and agree within $\sim$50\% at $\mapp=27$ (though worsening towards low survey area). Our predictions diverge more significantly at a survey area of 100 arcmin$^2$, where \citet{Ucci2020} have the fewest independent volumes in their simulation. Also, note that differences in the underlying UVLF can strongly affect the strength of Poisson noise; when Poisson noise begins to dominate, our results should not be too closely compared with those of \citet{Ucci2020}.

Our results (red solid lines) are systematically lower than those from \citet{Bhowmick2020} (dotted blue lines). However, our predictions remain within $\sim$25\% of each other except at smaller survey areas. At $\mapp = 30$, our predictions diverge at an area of $\sim10$ arcmin$^2$. At $\mapp=27$, our predictions diverge for survey areas where Poisson noise begins to dominate.

Numerical simulations have the benefit of being able to capture the non-linear bias of haloes. This effect, along with differences in Poisson noise from differing mass functions, could help explain the discrepancy between our predictions and those of the simulations at small survey areas.

In comparison to estimates with numerical simulations, the principal benefit of our model is its flexibility. 
We can test our model with any mass function or star formation and feedback prescription. Simulations must also subtract their intrinsic Poisson noise (which is not known perfectly) to estimate the cosmic variance, while analytic models can easily separate the two effects. 
Finally, we cover a wider range of redshifts ($z=5$--15) and magnitudes ($\mapp = 22$--38), and we can study larger volumes than simulations.

Our results agree quite well with those of \citet{Ucci2020}, over the range to which we can compare, especially at faint luminosities. For bright sources, our estimates are slightly below theirs, but the discrepancy is comparable to the apparent uncertainty in the CMF ($\sim$25\%; see Section~\ref{sec:uncertainty}). We agree reasonably well with \citet{Bhowmick2020} on large scales as well.

%%%%%%%%%%%%%%%%%%%%%%%%%%%%%%%%%%%%%%%%%%%%%%%%%%%%%%%%%%%%%%%%
%%%%%%%%%%%%%%%%%%%%%%%%%%%%%%%%%%%%%%%%%%%%%%%%%%%%%%%%%%%%%%%%

%%%%%%%%%%%%%%%%%%%%%%%%%%%%%%%%%%%%%%%%%%%%%%%%%%%%%%%%%%%%%%%%
%%%%%%%%%%%%%%%%%%%%%%%%%%%%%%%%%%%%%%%%%%%%%%%%%%%%%%%%%%%%%%%%
\section{Impact on Future Surveys}\label{sec:survey}

\begin{figure*}
    \centering
    \includegraphics[width=7.0in]{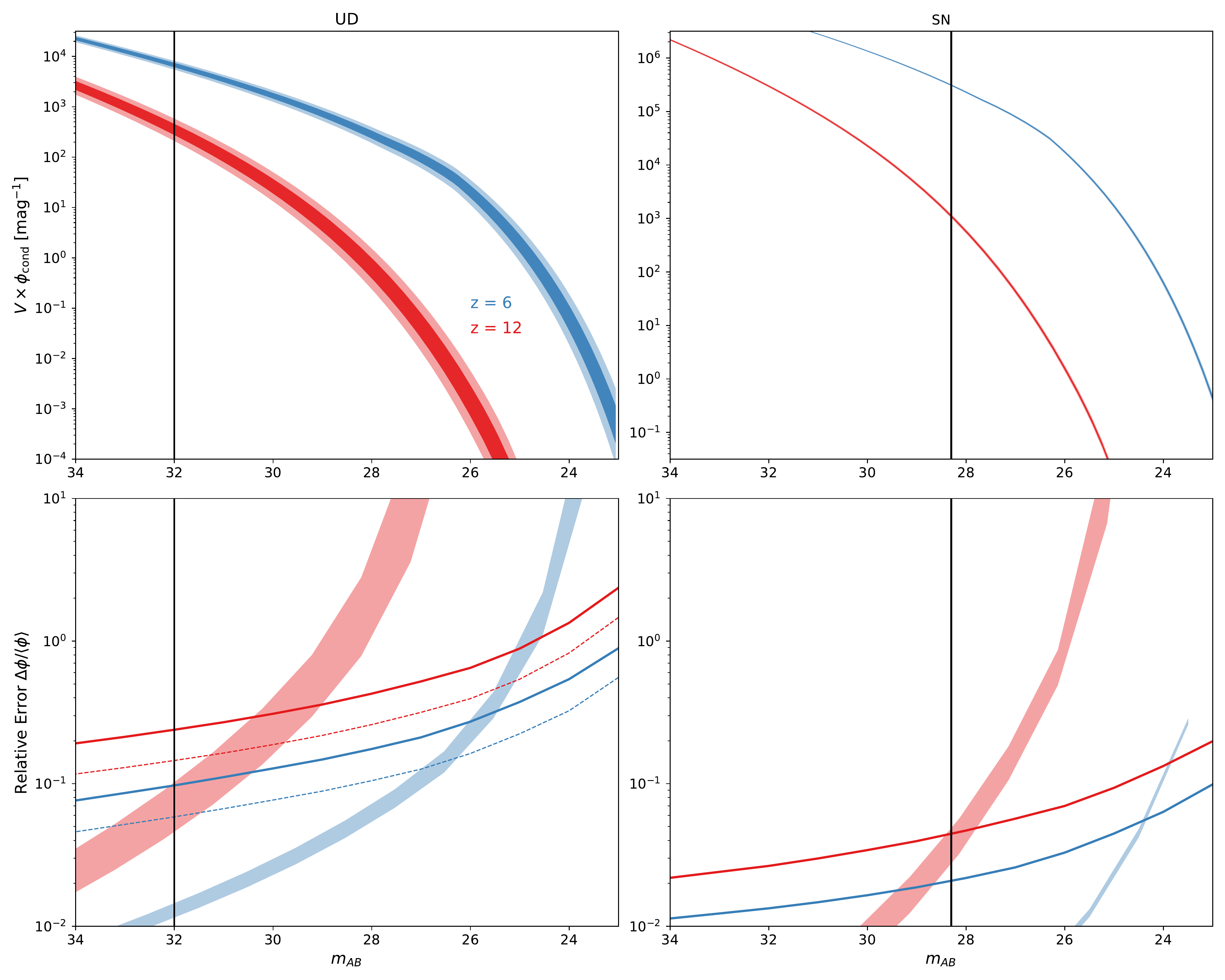}
    \caption{The effects of cosmic variance on the UVLFs of two high redshift surveys (UD and SN). In the upper panel, the width of the curves represents the 1$\sigma$ and 2$\sigma$ (inner and outer shading) ranges of intrinsic UVLFs that could be found in that survey's volume. In the lower panels, the lines show $\varepsilon_{cv}$ (upper set of lines are at $z = 12$, lower set at $z = 6$). If the UD survey is broken up into four independent pointings of JWST, each smaller sub-pointing has a higher variance than a large mosaic, but they may be combined for an overall reduction in measuring the average UVLF. This improvement is represented with the thin dashed lines. The vertical black lines are the magnitude limits of the surveys. The shaded band provides an estimate of Poisson shot noise. Cosmic variance acts as a noise floor for measuring the average UVLF, bounded on the faint end by the magnitude limit, and on the bright end by Poisson noise, except for the SN survey, which is entirely dominated by Poisson noise at high redshift.}
    \label{fig:money}
\end{figure*}

Cosmic variance will provide an unavoidable source of error for next generation telescopes, especially at the highest redshifts. It will dominate over Poisson noise for all but the brightest sources, and it is not easily avoided with deeper observations. Instead, it can only be minimized by probing larger volumes (at the cost of missing the more numerous faint sources) or by splitting up surveys into multiple independent pointings (at the cost of missing large-scale structure and making clustering measurements more difficult).

In this section, we perform a case study of the effects of cosmic variance on two upcoming instruments, JWST and the Roman Space Telescope. We consider two potential high redshift surveys: a JWST ultradeep (UD) survey \citep[following][]{Mason2015} along with a much wider-field Roman Space Telescope survey (similar to their planned supernova survey, and which we refer to as our SN survey). The UD survey has a detection limit of $\mapp\approx32.0$ and survey area of $A=40~$arcmin$^2$, while the SN survey has a detection limit of $\mapp\approx28.3$ and a survey area of  $A=9~$deg$^2$.
To begin, we assume that both are performed over a single contiguous area, requiring at minimum four and $30$ separate pointings (neglecting overlap between the pointings).

Additionally, we present a method of fitting an average UVLF to data from the UD and SN surveys simultaneously. Our method makes use of our model effectively as a ``prior'' on the cosmic variance in each survey field. The fitting process also accounts for the difference in shape between the local UVLF in each field and the average UVLF that we wish to fit.

%%%%%%%%%%%%%%%%%%%%%%%%%%%%%%%%%%%%%%%%%%%%%%%%%%%%%%%%
\subsection{Effects of cosmic variance on UD and SN surveys}

We show the effects of cosmic variance on the UVLF of the UD and SN surveys in Figure~\ref{fig:money}. The upper panels show the 1 and 2$\sigma$ fluctuations of the UVLF at $z = 6$ and $z = 12$. The lower panels show $\varepsilon_{cv}$ for these surveys (lines) and Poisson shot noise (shaded bands\footnote{Poisson shot noise is model-dependent, so we represent it as a band that encompasses the predictions from the three different feedback prescriptions described in section \ref{sec:FRSF} as well as the variety of number counts predicted from cosmic variance itself.}).
The vertical lines denote the magnitude limit of the surveys.

A given survey has access to the UVLF over a limited magnitude range, bound on the faint side by the magnitude limit and on the bright side by Poisson noise. In between, the noise floor of cosmic variance determines the maximum accuracy one can achieve in measuring the average UVLF over the accessible magnitude range if using just the one survey.

Splitting up a survey into independent pointings can improve the measurement of the average UVLF. While each individual pointing has higher cosmic variance than a large mosaic, they may be combined, which results in a reduction by the square root of the number of fields. The effect of splitting the UD survey into 4 pointings is represented by the dashed lines in Figure~\ref{fig:money}. \citet{Robertson2010} found a similar improvement in cosmic variance when splitting surveys into independent volumes. See Section \ref{sec:surveyDesign} for more details on the benefits/drawbacks of splitting up surveys

When interpreting survey results, it is crucial to note that cosmic variance is correlated across all magnitudes. If a survey probes a $1\sigma$ underdense region, the expected number counts in \textit{each} magnitude bin will be below the average by $1\sigma$. In contrast, Poisson noise is uncorrelated between each magnitude bin, depending only on the expected number of sources in that bin.

%%%%%%%%%%%%%%%%%%%%%%%%%%%%%%%%%%%%%%%%%%%%%%%%%%%%%%%%
\subsection{Measuring the average UVLF}\label{sec:lumFunMeasure}

Here we introduce a method to account for cosmic variance in measuring the  \textit{average} UVLF of the Universe given data from multiple independent survey volumes.
As an example of this method, we simulate mock UD and SN surveys of the UVLF and fit a model that extracts the average UVLF parameters; we then repeat this many times and compare those fits to the ``true'' parameters predicted by our model.

In this section, we model the average UVLF as a modified Schechter function:
\begin{equation}\label{eq:Schechter}
    \phi(L)dL=\frac{\phi^*}{L^*}\left(\frac{L}{L^*}\right)^\alpha e^{-(L/L^*)^\Gamma} dL,
\end{equation}
where $\phi(L)dL$ is the number density of galaxies with luminosities in the range 
$(L,L+dL)$, $\phi^*$ is a normalization constant, $L^*$ is the location of the exponential cutoff, $\alpha$ is the faint end slope, and $\Gamma$ is a parameter that governs the strength of the exponential cutoff. $\Gamma=1$ corresponds to a normal Schechter function. Our models are fit best with $\Gamma=0.5$, so we will use that value for this paper. We note that our use of $\Gamma=0.5$ predicts a higher number of bright galaxies than a normal Schechter function. This effect is reminiscent of recent studies of very high-$z$ surveys, which have found that the UVLF can be better fit by a double power-law due to an excess of bright galaxies \citep{Bowler2014,Bowler2020}. We do not use a double power law as our models are better fit by the modified Schechter function.

We explore four possible methods to measure the average UVLF of the Universe.
\begin{enumerate}
    \item \NoCV: We assume cosmic variance does not exist. Every region of the Universe has the exact same underlying UVLF, so Poisson noise is the only source of error.
    \item \Incorrect: Cosmic variance exists, but we fit the average UVLF without attempting to account for it.
    \item \Standard: We fit for the average UVLF using a common method to account for cosmic variance
    \item \Full: Our fiducial method. We fit for the average UVLF parameters using the conditional UVLF developed in this paper\footnote{A similar method is implemented in \citet{Livermore2017}; they consider cosmic variance in lensed surveys, and construct a conditional luminosity function using cosmic variance estimates from \citet{Robertson2014}.} (eq.~\ref{eq:CLF}).
    
\end{enumerate}

\begin{figure}
    \centering
    \includegraphics[width=0.5\textwidth]{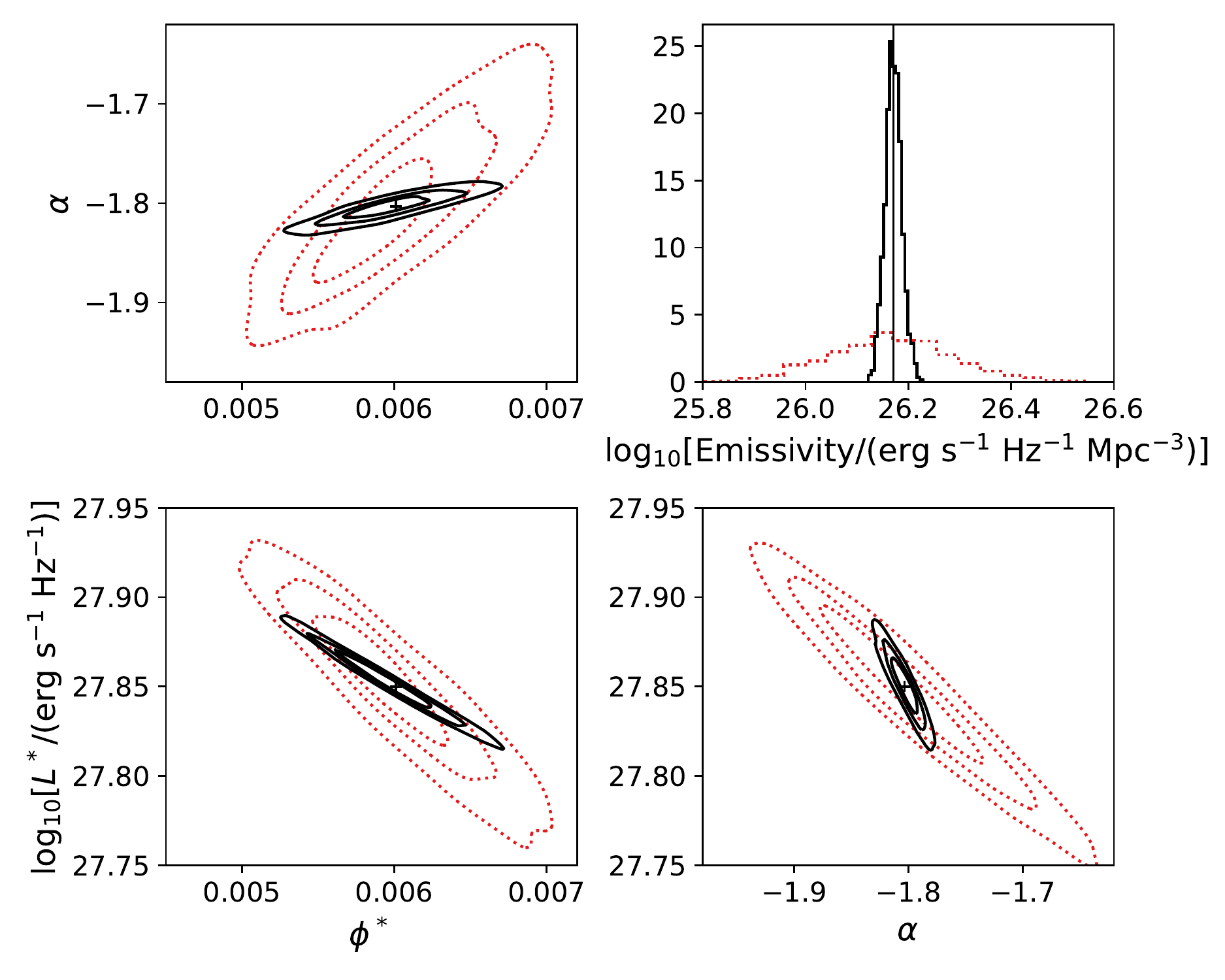}
    \caption{
    Recoveredluminosity functionparameters and uncertainties in the \NoCV~(\textit{solid contours}) and \Incorrect~(\textit{red dotted contours}) methods. The contours represent the distribution of best fit average UVLF parameters (see eq.~\ref{eq:Schechter}) for \pnum simulated pairs of UD and SN surveys at $z$ = 9. Cosmic variance adds a large amount of uncertainty to the determination of the ``true'' parameters (\textit{black crosses}) if not treated properly in the fitting technique. The top-right panel shows the emissivity of each fit (integrated down to $\massh_{\rm min}$), with the ``true'' average emissivity shown as the vertical line. The contours in this and all other figures are equally spaced between zero and the peak value of each normalized distribution.}
    \label{fig:plantain_z9_NoCV_Naive}
\end{figure}

\begin{figure}
    \centering
    \includegraphics[width=0.5\textwidth]{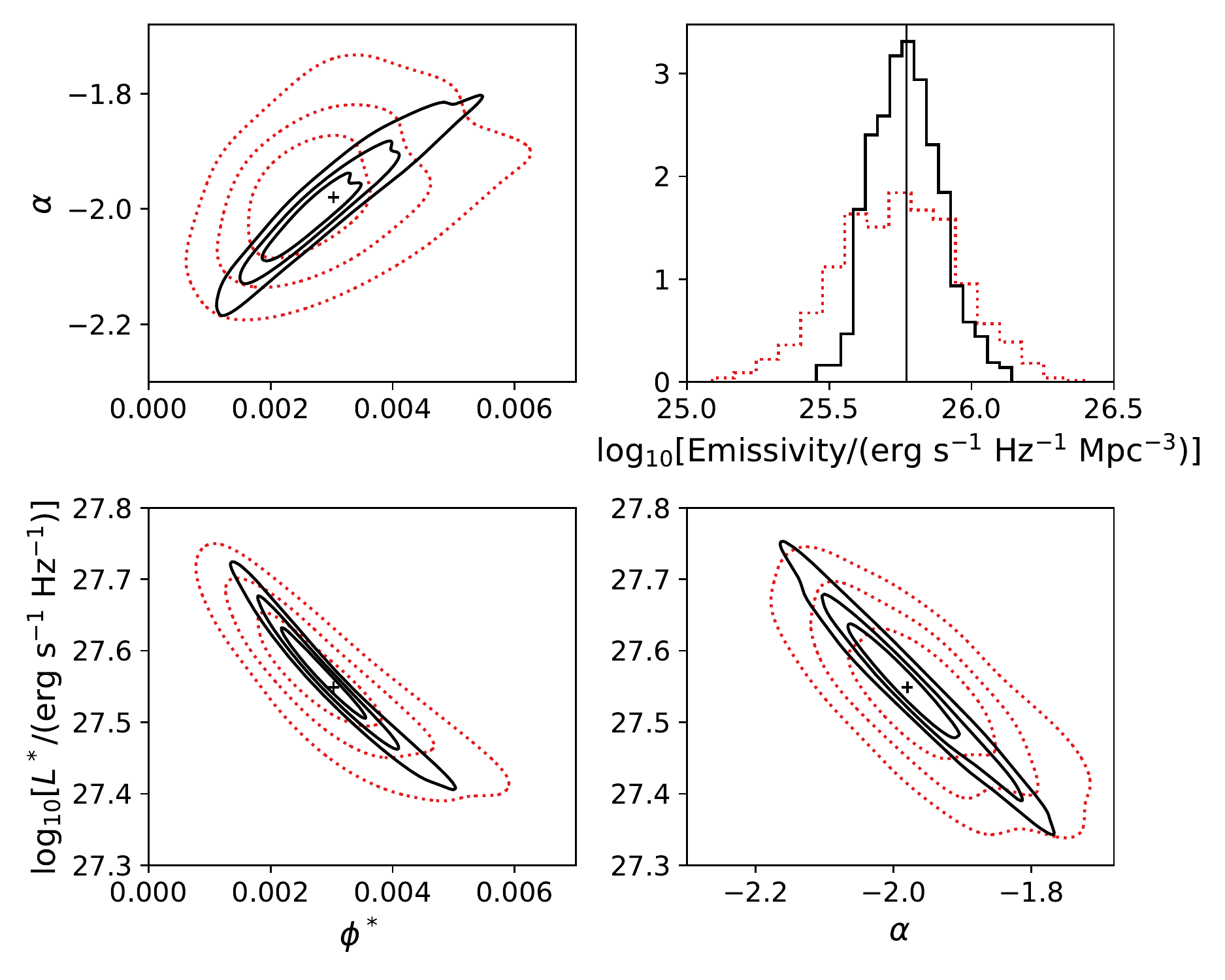}
    \caption{The same as Figure~\ref{fig:plantain_z9_NoCV_Naive} but at $z=12$.}
    \label{fig:plantain_z12_NoCV_Naive}
\end{figure}

The \NoCV~method assumes (unrealistically!) that cosmic variance does not exist. We simulate galaxy counts for the UD and SN surveys by drawing from the average UVLF that our model predicts, adding Poisson noise, and then fitting equation~(\ref{eq:Schechter}) to the combined mock data\footnote{We assume in this paper that the UD and SN surveys are perfect, in that they detect every galaxy and are able to accurately place each source in a magnitude bin of width $\Delta \mapp = 0.5$ and a redshift bin of $\Delta z = 1$. These are clearly not all accurate assumptions, especially the first one, but this treatment may be taken as a best possible scenario.}. 
The solid curves in Figures~\ref{fig:plantain_z9_NoCV_Naive} and \ref{fig:plantain_z12_NoCV_Naive} show the probability density functions (pdfs) of the best fits of the UVLF to \pnum sets of simulated data with \textit{no} cosmic variance (for $z = 9$ and 12, \textit{solid lines}). Unsurprisingly, this method recovers the ``true'' values (black crosses) of the average UVLF parameters well, as the SN probes the bright end and the UD the faint end, with some overlap between. Of course, cosmic variance \textit{does} exist; this method is only to be used as a comparison to our more realistic scenarios.

For the other three methods we use our model to simulate data for each survey, including cosmic variance. We first draw from the distribution of possible density environments for the UD survey $p(\delta_b|\eulrad,z)$ (see Appendix~\ref{app:EulCorr}) and then use equation~(\ref{eq:CLF}) to generate the UVLF for that survey\footnote{For $\langle \phi(\mapp) \rangle$ in equation~(\ref{eq:CLF}), we use the average UVLF predicted by our model, fit by equation~(\ref{eq:Schechter}) to obtain the ``true'' parameters.}. We then calculate the expected number of galaxies in each magnitude bin and apply Poisson shot noise. We repeat these steps for the SN survey. We then repeat this process \pnum times to generate many possible pairs of surveys.

\begin{figure}
    \centering
    \includegraphics[width=0.5\textwidth]{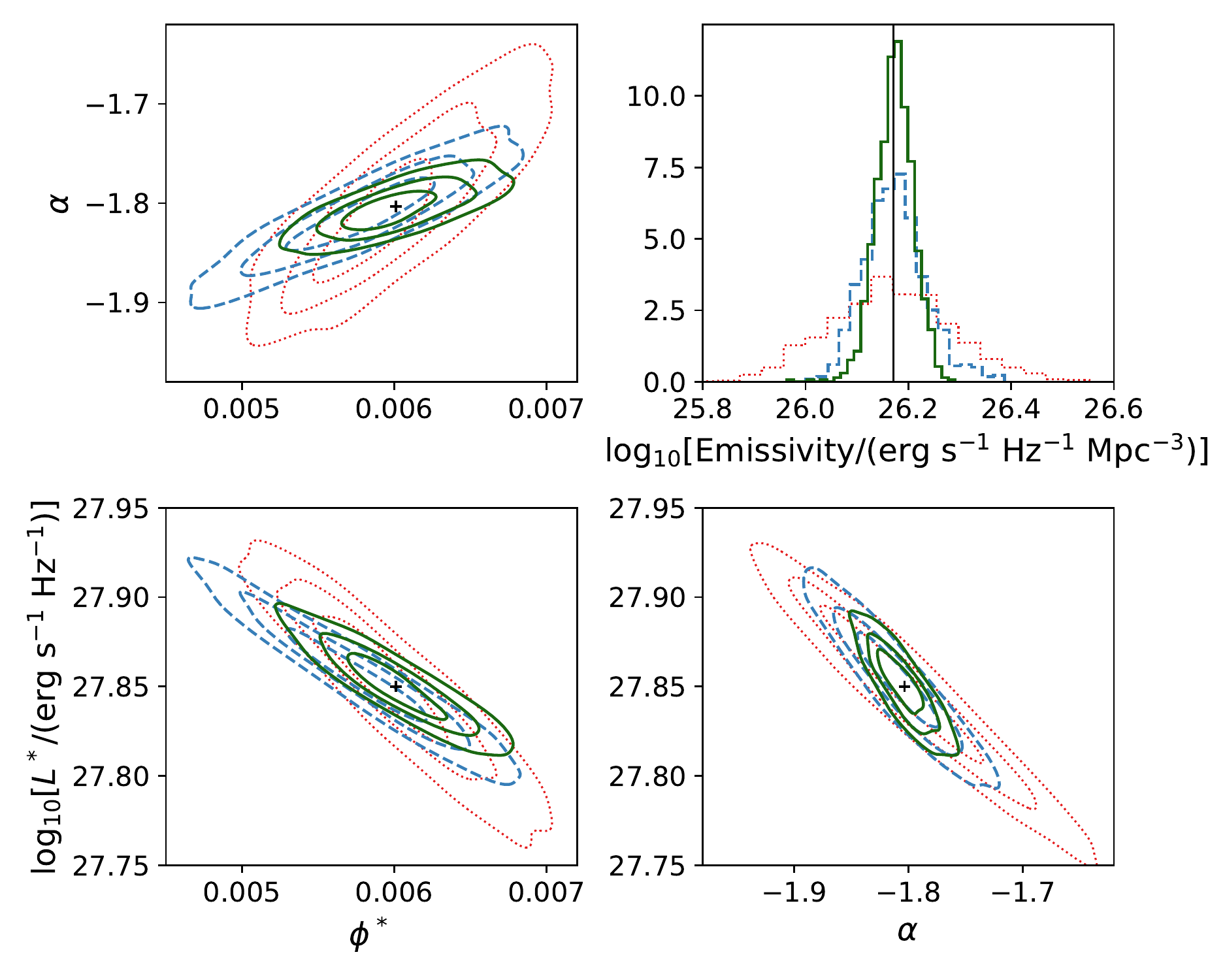}
    \caption{
    Recovered luminosity function parameters and uncertainties in the \Full~(\textit{green solid contours}), \Standard~(\textit{blue dashed contours}), and \Incorrect~(\textit{red dotted contours}) methods. The contours represent the distribution of best fit average UVLF parameters (see eq.~\ref{eq:Schechter}) for \pnum simulated pairs of UD and SN surveys at $z$ = 9. The \Full~and \Standard~methods are significant improvements over the \Incorrect~method, though the \Full~method does the best job recovering the ``true'' parameters (\textit{black crosses}). The \Standard~method is also slightly biased towards recovering a high $L^*$, low $\phi^*$, and steeper $\alpha$ in this case.
    \label{fig:plantain_z9_Full_BWNS}}
\end{figure}

\begin{figure}
    \centering
    \includegraphics[width=0.5\textwidth]{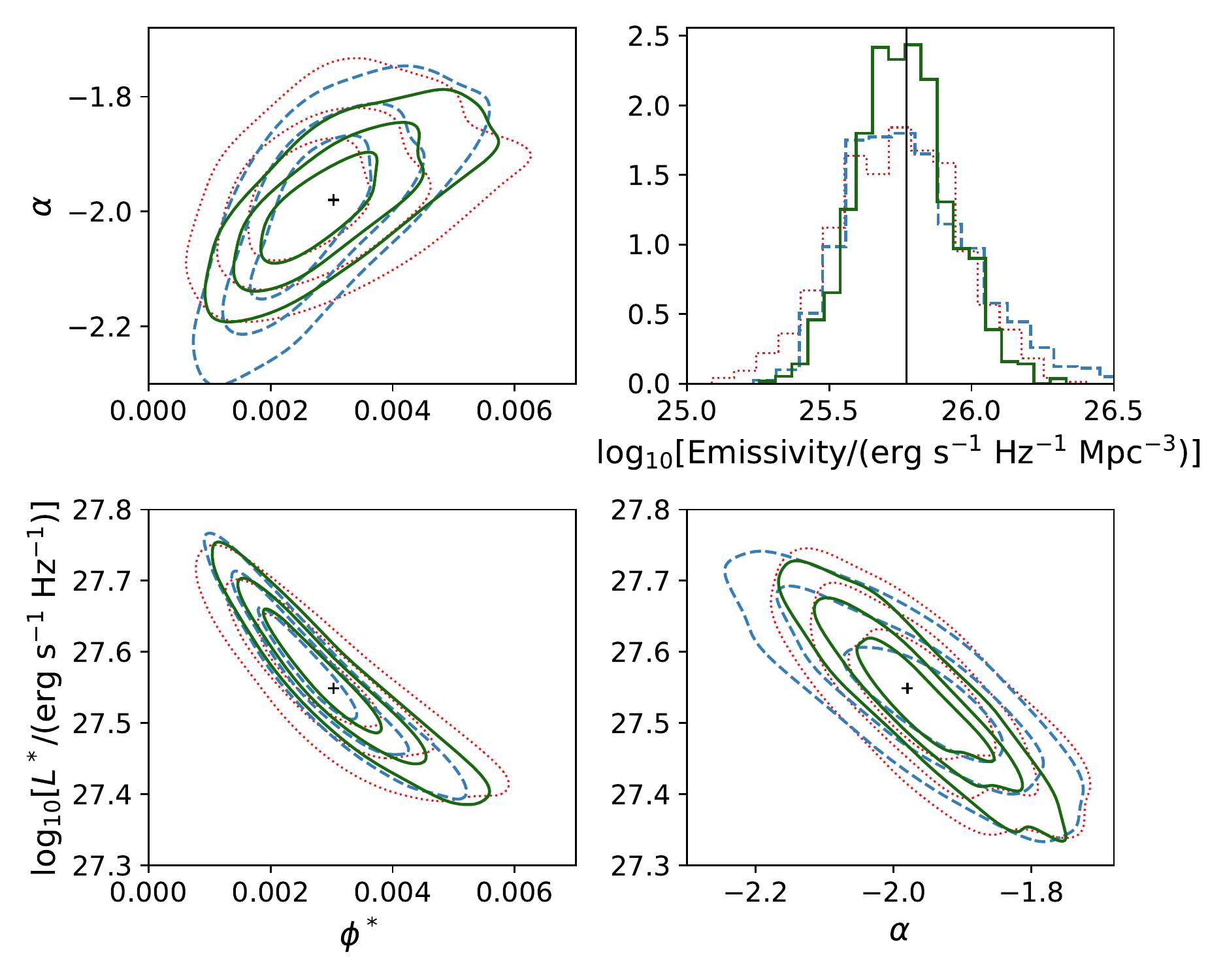}
    \caption{The same as Figure~\ref{fig:plantain_z9_Full_BWNS} but at $z=12$. The difference between the \Full, \Standard, and \Incorrect~models is less pronounced at $z=12$, though the \Full~method still performs best, and with the least amount of bias.}
    \label{fig:plantain_z12_Full_BWNS}
\end{figure}

In the \Incorrect~method, we simply joint fit equation~(\ref{eq:Schechter}) to the \pnum UD+SN mock data pairs with no attempt to correct for cosmic variance. The red dotted lines in Figures~\ref{fig:plantain_z9_NoCV_Naive}--\ref{fig:plantain_z12_Full_BWNS} show the resulting best fit pdf for the average UVLF. The recovered parameter range is far wider than the \NoCV~method because the \Incorrect~method completely ignores the effects of cosmic variance; the measured luminosity functions in the two surveys are \emph{not} the same so cannot easily be reconciled by a single fit.

The \Standard~method, originally developed by \citet{Sandage1979} and used by e.g. \citet{Efstathiou1988} and \citet{Bouwens2015}, fits a universal \textit{shape} to the UVLF, ignoring the field-to-field normalization. Then, the normalization is fixed at the end to reproduce the correct total number of galaxies across all surveys. Using this method, we fit to the mock data with cosmic variance. The blue dashed lines in Figures~\ref{fig:plantain_z9_Full_BWNS} and \ref{fig:plantain_z12_Full_BWNS} show the pdfs of the best fit parameters for the average UVLF. This method recovers the average UVLF parameters much more accurately than the \Incorrect~method.

While the \Standard~method is relatively robust to cosmic variance, it does not take into account any changes in the shape of the UVLF due to environment. Additionally, it does not incorporate any information about expected levels of cosmic variance, and it can produce biased results, as seen in this example by its systematic underprediction of the values of $\phi^*$ and $\alpha$ (most noticeably in Fig.~\ref{fig:plantain_z9_Full_BWNS}, upper-left panel).

Finally, in the \Full~method, we fit equation~(\ref{eq:CLF}) (with the modified Schecter function in eq.~\ref{eq:Schechter} as $\langle \phi(\mapp) \rangle$) simultaneously to each of the pairs of mock surveys, allowing for different values of $\delta_b$ for each survey. The green solid lines in Figures~\ref{fig:plantain_z9_Full_BWNS} and \ref{fig:plantain_z12_Full_BWNS} show the pdfs of best fit parameters for the average UVLF. Unsurprisingly (because we are fitting with the same function used to generate the mock data), the ``true'' parameters are recovered better than with the \Standard~method.

The upper right panels of Figures~\ref{fig:plantain_z9_NoCV_Naive}--\ref{fig:plantain_z12_Full_BWNS} show the total emissivity of the Universe as inferred from the parameters of the best fit (integrating down to $\massh_{\rm min}$), compared to the ``true'' average emmisivity that our model predicts (vertical line). The \Full~method does a slightly better job at recovering the average emissivity of the Universe compared to the \Standard~method, and both do much better than the \Incorrect~method.

While it is certainly to be expected that the \Full~method performs better than the \Standard~method in our calculations (given that we use our model to generate the mock data \textit{and} to fit to the data), the \Full~method still has benefits. First, it provides estimates of the dark matter overdensity $\delta_b$ for each survey field, while the \Standard~method by design throws out field-to-field variance information. Thus, the \Full~method can be used to test our understanding of cosmic variance, because it effectively has a prior on the level of cosmic variance allowed. It penalizes very high field-to-field variance, unlike the \Standard~method that effectively uses a flat prior on the amount of cosmic variance that is allowed during fitting. If real data were fit with the \Full~and \Standard~methods, and the \Standard~method provided a better fit, that would indicate that our understanding of cosmic variance is flawed. We could use our model to investigate \textit{where} and \textit{why} our understanding of cosmic variance breaks down in terms of our physically motivated inputs.

One could investigate the time evolution of UVLF parameters after determining the best fit values at a variety of redshifts. However, this experiment would need to be done with care, as Figures~\ref{fig:plantain_z9_NoCV_Naive}--\ref{fig:plantain_z12_Full_BWNS} show that the UVLF parameters are highly correlated.
Thus, their time evolution must be fit jointly and with a good estimation of their covariance.
That covariance can depend strongly on the treatment of cosmic variance.

%%%%%%%%%%%%%%%%%%%%%%%%%%%%%%%%%%%%%%%%%%%%%%%%%%%%%%%%
\subsection{The benefits of multiple surveys}

\begin{figure}
    \centering
    \includegraphics[width=0.5\textwidth]{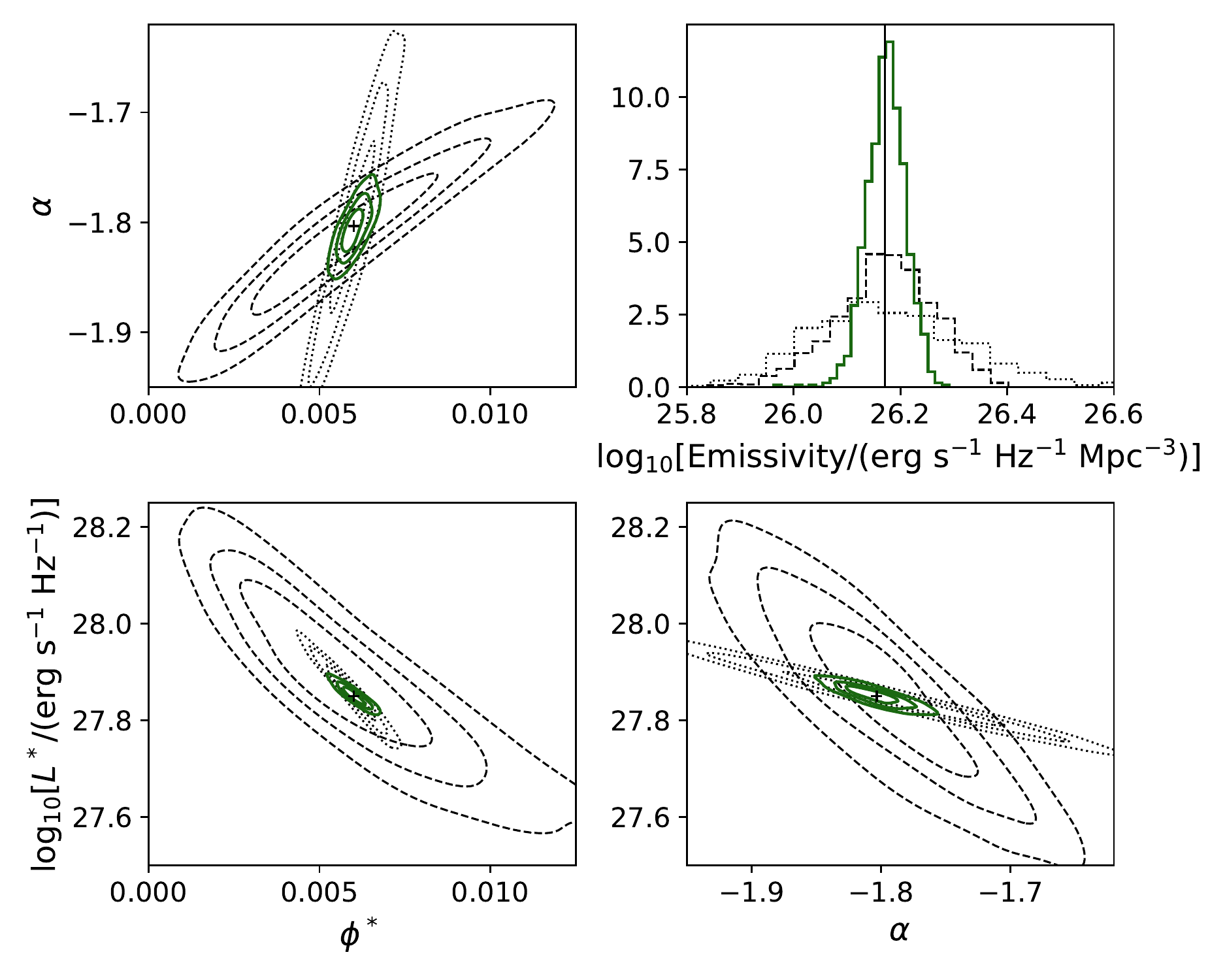}
    \caption{The range of parameters obtained with the SN survey alone (\textit{dotted contours}), the UD survey alone (\textit{dashed contours}), and with both simultaneously fit (\textit{green solid contours}, same as those in  Figure~\ref{fig:plantain_z9_Full_BWNS}). The upper right panel shows the distribution of emissivities calculated from these distributions.}
    \label{fig:plantain_z9_OneSurvey}
\end{figure}

Next we consider the importance of measuring the UVLF with multiple complementary surveys. Figure~\ref{fig:plantain_z9_OneSurvey} shows the range of best fit parameters for the $z=9$ UVLF when fitting to SN survey data alone (\textit{dotted contours}), the UD survey alone (\textit{dashed contours}), and with both simultaneously fit (\textit{solid contours}, identical to those in  Figure~\ref{fig:plantain_z9_Full_BWNS}). The SN survey alone provides good constraints on $\phi^*$ and $L^*$, but the faint end slope $\alpha$ is constrained better by the joint fit than either survey alone.

\begin{figure}
    \centering
    \includegraphics[width=0.5\textwidth]{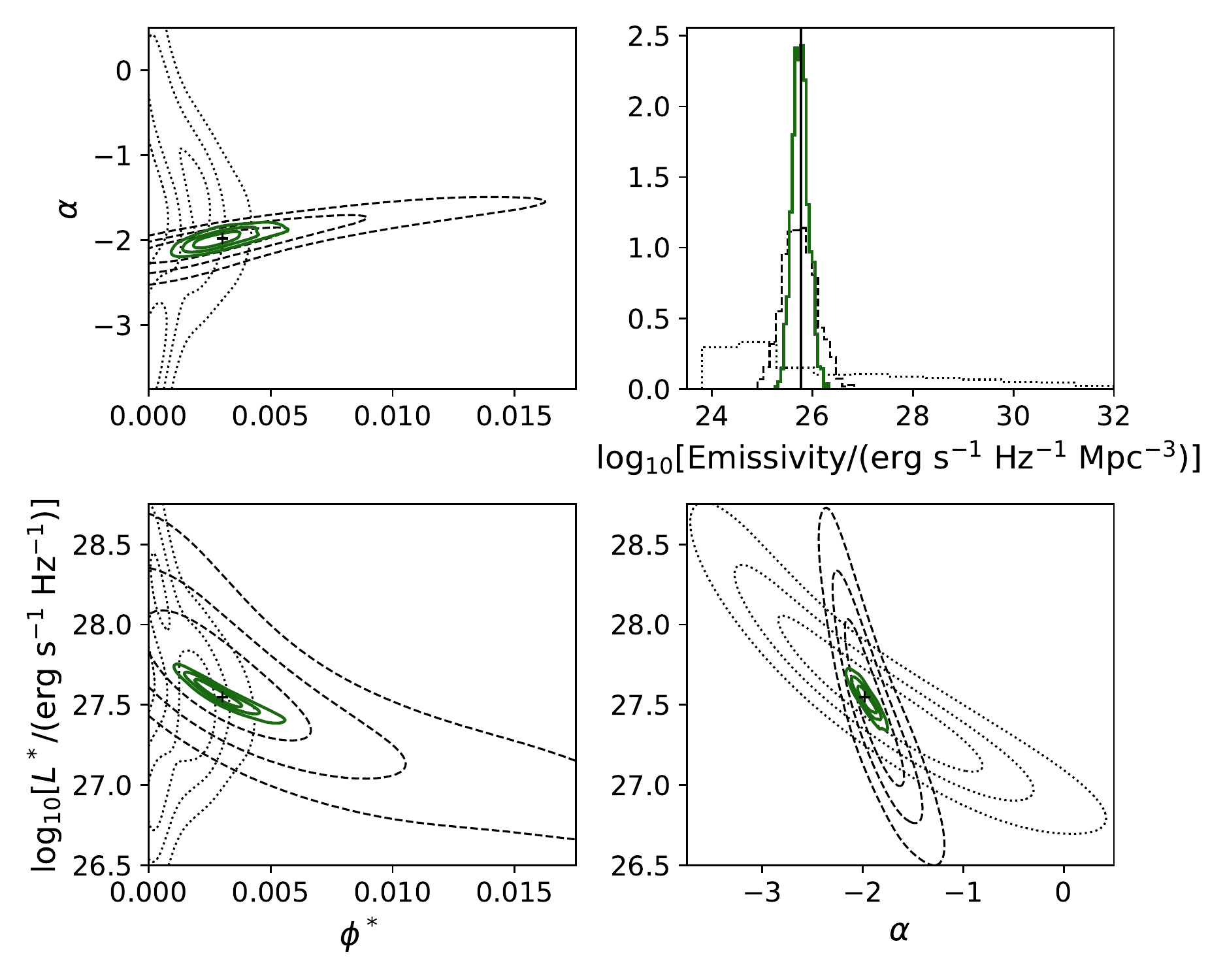}
    \caption{Same as Figure~\ref{fig:plantain_z9_OneSurvey} but at $z=12$. Each individual survey has much lower constraining power alone.}
    \label{fig:plantain_z12_OneSurvey}
\end{figure}

At $z=12$, shown in Figure~\ref{fig:plantain_z12_OneSurvey}, the combination of these two surveys is even more crucial, as neither survey can provide good constraints on any parameter by itself. 

We also investigated the effects of splitting up the UD survey into four independent pointings and re-running the \Full~and \Standard~methods. This method gives a significantly better determination of the average number density of very faint sources. However, it only results in a slightly better determination of the average UVLF parameters, because the faint-end slope $\alpha$ is not very sensitive to cosmic variance, and the SN survey dominates the constraints of $\phi^*$ and $L^*$.

We see that tiered surveys, including both wide and deep strategies, will be essential for providing an accurate census of the high-$z$ galaxy population.

%%%%%%%%%%%%%%%%%%%%%%%%%%%%%%%%%%%%%%%%%%%%%%%%%%%%%%%%
\subsection{Time allocation and survey design strategies}\label{sec:surveyDesign}

One important use for our results is to identify survey design strategies that result in the best constraints on the average UVLF parameters (Figs.~\ref{fig:plantain_z9_Full_BWNS} and \ref{fig:plantain_z12_Full_BWNS}). 
One could use simulations of our model to optimize the design given constraints on telescope time, survey depth, and area, but here we provide a strategy for a good initial guess.

Given a single magnitude bin and an error requirement $\epsilon_{\rm req}$ in measuring the average number counts in that bin, there is a minimum survey area below which cosmic variance will exceed the error requirement. For example, say we wish to design a survey that measures the average UVLF at $z=9$ (and $\Delta z = 1$) at apparent magnitudes of 30 and 26 with contributions from cosmic variance at those magnitudes below 15\% and 10\%, respectively. Reading off Figure~\ref{fig:powerfit} or using our python package \pakidge, we find that these would require $\sim$300 arcmin$^2$ and $\sim$2.8 deg$^2$, respectively.

Alternatively, if we were willing to split each survey into independent pointings, we could satisfy the same error requirements with four $\sim$3 arcmin$^2$ surveys down to $\mapp = 30$ and nine $\sim$0.11 deg$^2$ surveys down to $\mapp = 26$. This observing plan requires $\sim$25$\times$ less telescope time for the deep survey and 
$\sim$3$\times$ less telescope time for the wide-field survey. Splitting up surveys is an especially efficient way to mitigate cosmic variance for narrow surveys because the curves in Fig.~\ref{fig:powerfit} are flattest at small survey area, so there is little penalty for moving to even narrower independent pointings. However, we do note that our model does not include nonlinear clustering that may become more important in such narrow survey fields.

Unfortunately, splitting up a survey into smaller and smaller sub-pointings is not without its drawbacks. Large mosaics can be used to measure clustering of galaxies; splitting up a survey leaves many spatial scales inaccessible, and clustering is typically more difficult to measure in the radial direction. Also, multiple small fields of view can miss interesting large-scale structures such as proto-clusters. Splitting surveys also increases observing overhead and survey design complexity. An efficient compromise would be a tiered approach: the majority of a survey's area is in one contiguous location, while a smaller fraction is split into a few independent pointings to calibrate for cosmic variance.

%%%%%%%%%%%%%%%%%%%%%%%%%%%%%%%%%%%%%%%%%%%%%%%%%%%%%%%%%%%%%%%%
%%%%%%%%%%%%%%%%%%%%%%%%%%%%%%%%%%%%%%%%%%%%%%%%%%%%%%%%%%%%%%%%

%%%%%%%%%%%%%%%%%%%%%%%%%%%%%%%%%%%%%%%%%%%%%%%%%%%%%%%%%%%%%%%%
%%%%%%%%%%%%%%%%%%%%%%%%%%%%%%%%%%%%%%%%%%%%%%%%%%%%%%%%%%%%%%%%
\section{Conclusions}\label{sec:conclusions}

Cosmic variance will be an unavoidable source of error for next generation telescopes when measuring average properties of the Universe, especially at higher redshifts. Cosmic variance will dominate over Poisson noise for all but the brightest sources. This study integrates cosmic variance into the galaxy model developed in \citet{Furlanetto2017}.
We first consider how star formation rates vary with environment in the model.
Next, we construct a conditional UVLF and provide its linear approximation for a wide variety of survey parameters with the parameter $\varepsilon_{cv}$ via equation~(\ref{eq:CLF}). We then study what parts of our model are most important in determining $\varepsilon_{cv}$.
Finally, we propose a method for using these estimates as a prior on cosmic variance to improve fitting luminosity functions to high-$z$ data.

In our model, the choice of star formation and feedback prescriptions has little effect on the relative strength of cosmic variance, and haloes of fixed mass are similar in all environments. Therefore, the main driver of cosmic variance in the UVLF is cosmic variance in the underlying dark matter halo population. The halo mass function is also the main driver in the uncertainty in our model; a more accurate conditional mass function would allow for a better prediction of cosmic variance.

A simple dark matter halo bias function along with an average halo mass to UV luminosity relation can adequately describe the relative effects of cosmic variance in the UVLF, except for density excursions exceeding $\sim$2$\sigma$. In those regions, cosmic variance becomes non-Gaussian, and a full treatment is required.

We provide linear approximations of cosmic variance via $\varepsilon_{cv}$ in terms of apparent (rest-UV) AB magnitude, survey area, and redshift. This approximation may be easily applied to any average UVLF via equation~(\ref{eq:CLF}). 
We provide a public python package \pakidge~for easy access to our results. This package provides values of $\varepsilon_{cv}$ over a wide range of redshifts, magnitudes, survey areas, and redshift bin widths. It also includes two options for the conditional mass function used, which can be used as an estimate of the model uncertainty in the value of $\varepsilon_{cv}$ (see Data Availability section for more details).
We compare our results with cosmic variance predictions from simulations 
\citep{Bhowmick2020,Ucci2020} and find good general agreement except at the smallest survey volumes (where Poisson noise begins to dominate and non-linear halo bias could be significant), or at volumes that are sizable fractions of their simulations' box size.

We also present a method for using our model as a prior on cosmic variance when fitting a UVLF to galaxy survey data. This method can inform our understanding of cosmic variance while also improving the quality of and reducing the bias in fitting the UVLF. It allows us to quantify the gains from splitting surveys into independent pointings and combining independent observations. In particular, we have shown that the combination of a shallow wide survey and a deep narrow survey is essential for fully constraining the UVLF. We also show that splitting up a survey can be an effective way to reduce the effects of cosmic variance.

Our model treats galaxy formation in a very simple manner. The primary simplification is in modelling only the \textit{average} galaxy population in a given environment. We also ignore the effects of dust, mergers, scatter in the halo mass to UV luminosity relation, the evolution of the IMF, and the spatial distribution of star formation within a dark matter halo. Fortunately, these shortcomings pertain to (1) the details of star formation, which we have shown hardly affect the \textit{relative} cosmic variance results $\varepsilon_{cv}$; and (2) individual galaxies, which are likely averaged out (to an extent) when considering cosmic variance in an ensemble of galaxies.

An understanding of cosmic variance is essential for quantifying the uncertainty in future surveys with observatories like JWST and the Roman Space Telescope. We hope that our flexible model, and the method we have introduced to incorporate cosmic variance explicitly into fitting multiple fields, can offer better constraints not just on the galaxy luminosity function but also on cosmic variance itself.

%%%%%%%%%%%%%%%%%%%%%%%%%%%%%%%%%%%%%%%%%%%%%%%%%%%%%%%%%%%%%%%%
%%%%%%%%%%%%%%%%%%%%%%%%%%%%%%%%%%%%%%%%%%%%%%%%%%%%%%%%%%%%%%%%

%%%%%%%%%%%%%%%%%%%%%%%%%%%%%%%%%%%%%%%%%%%%%%%%%%%%%%%%%%%%%%%%
%%%%%%%%%%%%%%%%%%%%%%%%%%%%%%%%%%%%%%%%%%%%%%%%%%%%%%%%%%%%%%%%
\section*{Acknowledgements}

We thank Frederick B. Davies, Jordan Mirocha, Guochao Sun, Brant E. Robertson, and our reviewer for helpful conversations and suggestions.
We also would like to express our gratitude and thanks to Richard H. Mebane for lending knowledge and guidance throughout, Jon K. Zink for providing expertise on statistical methods, and John A. Trapp for help with publishing the \pakidge~package.

This work was supported by the National Science Foundation through award AST-1812458. In addition, this work was directly supported by the NASA Solar System Exploration Research Virtual Institute cooperative agreement number 80ARC017M0006. We also acknowledge a NASA contract supporting the ``WFIRST Extragalactic Potential Observations (EXPO) Science Investigation Team" (15-WFIRST15-0004), administered by GSFC. 

\textit{Software used:} This work makes use of iPython \citep{Perez2007} and the following Python packages: NumPy \citep{Walt2011}, SciPy \citep{Virtanen2020}, Matplotlib \citep{Hunter2007}, and pandas \citep{McKinney2010}.

%%%%%%%%%%%%%%%%%%%%%%%%%%%%%%%%%%%%%%%%%%%%%%%%%%%%%%%%%%%%%%%%
%%%%%%%%%%%%%%%%%%%%%%%%%%%%%%%%%%%%%%%%%%%%%%%%%%%%%%%%%%%%%%%%

%%%%%%%%%%%%%%%%%%%%%%%%%%%%%%%%%%%%%%%%%%%%%%%%%%
\section*{Data Availability}\label{sec:dataAvail}

We provide a public Python package \pakidge~that can be used to calculate $\varepsilon_{cv}$ for a wide range of survey parameters (redshift, apparent AB magnitude, survey area, and redshift bin width). This package may be installed in a Python environment via `\verb'pip install galcv'' and then imported via `\verb'import galcv''.
The code for this package and instruction for its installation and use may be found online at \textbf{https://github.com/adamtrapp/galcv}. Requests for bug fixes and suggestions for additions are welcome.



\bibliographystyle{mnras}
\bibliography{cv} % if your bibtex file is called example.bib

%\begin{thebibliography}{99}
%\end{thebibliography}

%%%%%%%%%%%%%%%%%%%%%%%%%%%%%%%%%%%%%%%%%%%%%%%%%%

%%%%%%%%%%%%%%%%% APPENDICES %%%%%%%%%%%%%%%%%%%%%

%%%%%%%%%%%%%%%%%%%%%%%%%%%%%%%%%%%%%%%%%%%%%%%%%%%%%%%%%%%%%%%%
%%%%%%%%%%%%%%%%%%%%%%%%%%%%%%%%%%%%%%%%%%%%%%%%%%%%%%%%%%%%%%%%
\appendix

%%%%%%%%%%%%%%%%%%%%%%%%%%%%%%%%%%%%%%%%%%%%%%%%%%%%%%%%
\section{The Eulerian Volume Correction}\label{app:EulCorr}

In this appendix we construct $p(\delta_b|\eulrad,z)d\delta_b$, the fraction of volume in the Universe with linear density between ($\delta_b$,$\delta_b+d\delta_b$) when averaged over the Eulerian scale $\eulrad$. This distribution is in contrast to $p(\delta_b|\lagrad,z)d\delta_b$, the fraction of \emph{mass} in the Universe with linear density between ($\delta_b$,$\delta_b+d\delta_b$) when averaging over the \emph{Lagrangian} scale $\lagrad$.

As described in section \ref{sec:densityfluc}, $p(\delta_b|\lagrad,z)$ is by definition equal to a zero-mean Gaussian with variance $\sigma^2(\massr,z)$, where $\massr$ is the mass of a region of radius $\lagrad$ and average density. Unfortunately, $p(\delta_b|\lagrad,z)$ considers a fixed mass scale $\lagrad$\footnote{$\lagrad$ is a mass (Lagrangian) scale because it is defined as the radius of a region of mass $\massr$ \textit{if} that region were at average density. In reality, regions of mass $\massr$ can have different physical volumes depending on their densities, as we will show in this Appendix.}, which corresponds to a density-dependent range of different volumes.

Let us choose one fixed Eulerian scale $\eulrad$. We consider that scale's corresponding Lagrangian radii
\begin{equation}\label{eq:constvol}
    \lagrad^3=\eulrad^3(1+\delta_r),
\end{equation}
where $\delta_r$ is the true, nonlinear density of the region. Following \cite{Mo1996}, the \textit{real} density of a region may be related to the linear density via the following approximation (assuming spherical collapse):
\begin{equation}\label{eq:realden}
\begin{aligned}
    \delta_b = &-1.35(1+\delta_r)^{-2/3}+0.78785(1+\delta_r)^{-0.58661} \\
    &-1.12431(1+\delta_r)^{-1/2}+1.68647.
\end{aligned}
\end{equation}

Inserting this value into equation~(\ref{eq:constvol}), we now have $\lagrad(\delta_b|\eulrad)$, a relation between linear density and Lagrangian radius at fixed Eulerian radius. We convert $\lagrad$ to $\sigma$ via $\sigma(\massr=4\pi/3~\bar\rho\lagrad^3)$ and convert $\delta_b$ to $\delta_0$ via the growth function $\delta_0 = \delta_b/F_g(z)$. That process provides $\sigma^2(\delta_0|\eulrad,z)$, a locus in ($\sigma^2$, $\delta_0$) space of constant Eulerian radius $\eulrad$.

With $\sigma^2(\delta_0|\eulrad,z)$, we can use the excursion set formalism to solve for $f_R(\sigma^2|\eulrad,z)$, the distribution of mass in the Universe that is associated with a region with $\sigma^2$ (and thus corresponding $\lagrad$ and $\delta_0$) at fixed $\eulrad$. The excursion set formalism describes a random walk in dark matter density $\delta_0$ as one averages over first a very large volume (small $\sigma)$, and then successively smaller volumes (larger $\sigma$) centered at a single point in space \citep{Bond1991,Lacey1993}. The distribution of random walks that first cross the barrier $\sigma^2(\delta_0|\eulrad,z)$ defines $f_R(\sigma^2|\eulrad,z)$.

$f_R(\sigma^2|\eulrad,z)$ has no analytic solution for an arbitrary barrier shape, so we approximate $\sigma^2(\delta_0|\eulrad,z)$ as a straight line,\footnote{We approximate $\sigma^2(\delta_0|\eulrad,z)$ as a line by fitting it to the barrier near where most trajectories cross the barrier: $\delta_0=0$.}
\begin{equation}\label{eq:barrier}
    B(\sigma^2|\eulrad,z)=B_0+B_1 \sigma^2,
\end{equation}
where $B$ is the density $\delta_0$, and $B_0$ and $B_1$ are fit parameters (corresponding to the $y$-intercept and slope, respectively).

Fortunately, the first-crossing distribution $f_R$ for a linear barrier in $(\sigma^2,\delta_0)$ space has been solved analytically by \citet{Sheth1998}:
\begin{equation}\label{eq:f_Lagr}
\begin{aligned}
    &f_R(\sigma^2,|\eulrad,z)\textrm{d}\sigma^2=\\
    &\frac{B(0|\eulrad,z)}{\sqrt{2\pi \sigma^2}}\textrm{exp}\left(-\frac{B^2(\sigma^2|\eulrad,z)}{2\sigma^2}\right)\frac{\textrm{d}\sigma^2}{\sigma^2}.
\end{aligned}
\end{equation}
This is an Inverse Gaussian distribution.

We convert $f_R(\sigma^2|\eulrad,z)$, a mass fraction distribution in $\sigma^2$, to $p(\delta_b|\eulrad,z)$, a volume fraction distribution in $\delta_b$, following equation (16) of \citet{Sheth1998}:
\begin{equation}\label{eq:Peul}
   p(\delta_b|\eulrad,z)\textrm{d}\delta_b=\frac{1}{(1+\delta_r)}f_R(\sigma^2|\eulrad,z)\textrm{d}\sigma^2.
\end{equation}
In principle, dividing by the non-linear function $(1+\delta_r)$ can result in a $p(\delta_b|\eulrad,z)$ that is not normalized. In practice, $p(\delta_b|\eulrad,z)$ remains normalized within 1\% for all cases we consider. 

\begin{figure}
    \centering
    \includegraphics[width=0.5\textwidth]{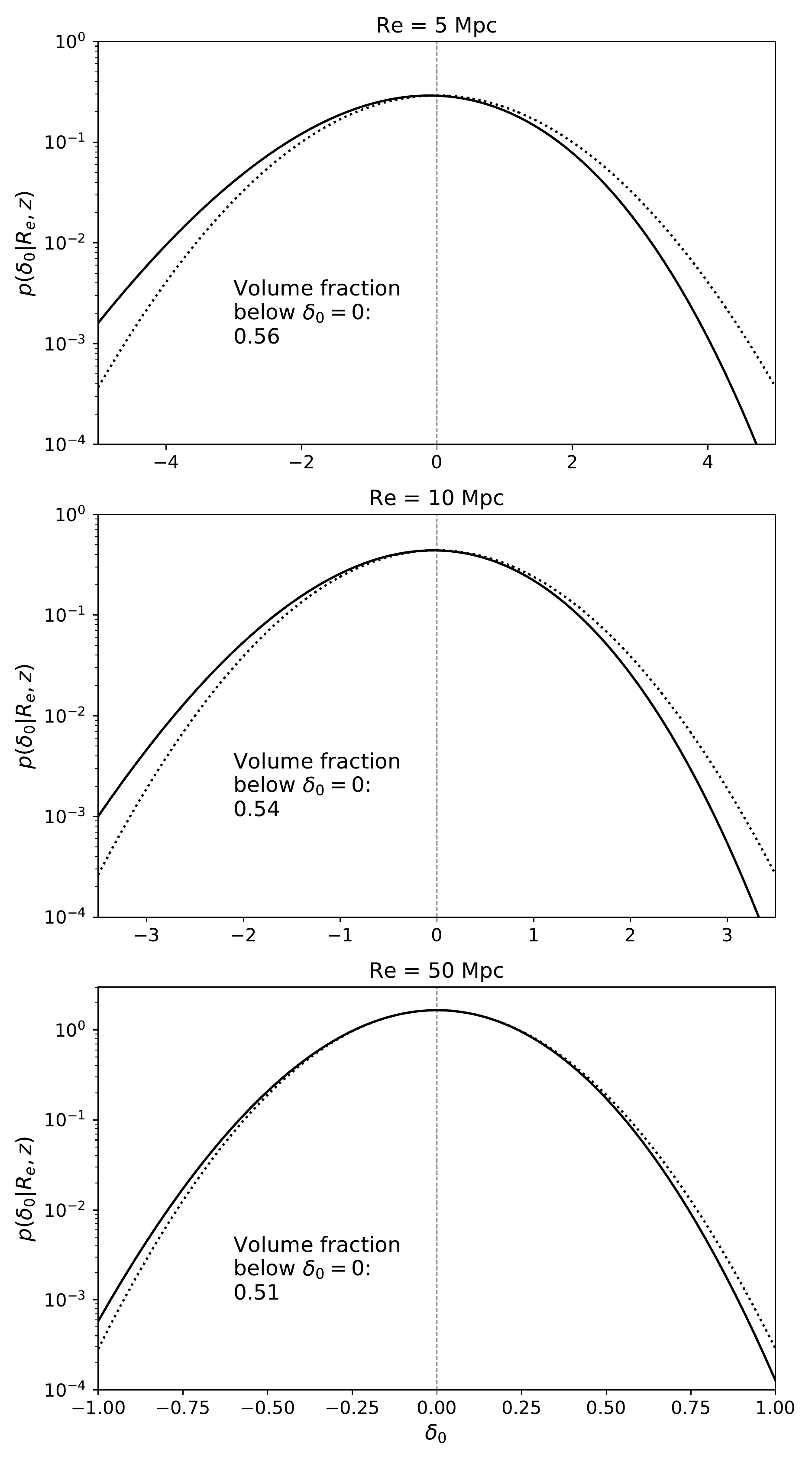}
    \caption{Effects of the Eulerian correction at $z = 9$. The \textit{solid} lines show $p(\delta_0|\eulrad,z)$, the distribution of linear densities at fixed scale (densities are extrapolated to $z = 0$ via the growth function). The \textit{dotted} lines show the Lagrangian distribution of densities $p(\delta_0|\lagrad,z)$: Gaussian distributions with standard deviation $\sigma(\eulrad,z)$. The volume of the Universe that is below average density at each scale is indicated in each panels.}
    \label{fig:denDist}
\end{figure}

For the range of redshifts and scales considered in this paper, $p(\delta_b|\eulrad,z)$ is near to a Gaussian with standard deviation $\sigma(\massr=4\pi/3~\bar\rho\eulrad^3)$. However, the distribution is skewed towards negative densities, resulting in a boost in the negative wing and suppression in the positive wing, an effect that is most significant for volumes with radii less than $\sim$10 Mpc (see Fig.~\ref{fig:denDist}).

At $z = 9$, we find that for regions with scales $\eulrad$ = 5, 10, and 50 Mpc, the fraction of volume in the Universe that is below average cosmic density is 56, 54, and 51\%, respectively. These fractions increase slightly at lower redshifts as underdense regions continue to expand relative to overdense regions. This result indicates that surveys will be slightly more likely to probe underdense regions.
Using different methods, \citet{Munoz2010} also found that surveys are more likely to probe an underdense region because of those regions' more rapid cosmic expansion.

%%%%%%%%%%%%%%%%%%%%%%%%%%%%%%%%%%%%%%%%%%%%%%%%%%%%%%%%%%%%%%%%
%%%%%%%%%%%%%%%%%%%%%%%%%%%%%%%%%%%%%%%%%%%%%%%%%%%%%%%%%%%%%%%%

%%%%%%%%%%%%%%%%%%%%%%%%%%%%%%%%%%%%%%%%%%%%%%%%%%

% Don't change these lines
\bsp	% typesetting comment
\label{lastpage}
\end{document}